\documentclass[10pt,journal,letterpaper,compsoc,twoside]{IEEEtran}
\usepackage{graphicx}   
\usepackage{listings}   
\usepackage[usenames]{color}

\ifCLASSOPTIONcompsoc
\else
\fi
\ifCLASSINFOpdf
\else
\fi
\newcommand{\ttt}[1]{\begin{small}\begin{tt}#1\end{tt}\end{small}}

\begin{document}
%
\title{Modular Workflow Engine for Distributed Services using Lightweight Java Clients}
%
%
%
%

\author{Ralf-Michael~Vetter,
        Werner~Lennartz,
        and~J\"org-Volker~Peetz
\IEEEcompsocitemizethanks{\IEEEcompsocthanksitem Dr. R.-M. Vetter is with the Department
of Mechanical Engineering, University of Applied Sciences Bonn-Rhein-Sieg, Sankt Augustin, Germany. E-mail: post@rmvetter.de
\IEEEcompsocthanksitem Dr. W. Lennartz is head of software development at BusinessActs Information and Data Systems,
Cologne, Germany. E-mail: see http://businessacts.de
\IEEEcompsocthanksitem Dr. J.-V. Peetz is with the Department of Simulation Engineering, 
Fraunhofer Institute for Algorithms and Scientific Computing (SCAI), Sankt Augustin, Germany. E-mail: see http://scai.fraunhofer.de}
}

%
%

\markboth{Preprint}%
{Vetter \MakeLowercase{\textit{et al.}}: Modular Workflow Engine for Distributed Services using Lightweight Java Clients}
%



\IEEEcompsoctitleabstractindextext{%
\begin{abstract}
In this article we introduce the concept and the first implementation of a lightweight 
client-server-framework as middleware for distributed computing. 
On the client side an installation without administrative rights or privileged ports can 
turn any computer into a worker node. Only a Java runtime environment and the JAR files 
comprising the workflow client are needed. To connect all clients to the engine one open 
server port is sufficient. The engine submits data to the clients and 
orchestrates their work by workflow descriptions from a central database. Clients request 
new task descriptions periodically, thus the system is robust against network failures. 
In the basic set-up, data up- and downloads are handled via HTTP communication with the server. 
The performance of the modular system could additionally be improved using dedicated file servers 
or distributed network file systems.

We demonstrate the design features of the proposed engine in real-world applications from mechanical 
engineering. We have used this system on a compute cluster in design-of-experiment studies,
parameter optimisations and robustness validations of finite element structures.
\end{abstract}

\begin{IEEEkeywords}
Java grid engine, workflow automation, distributed services, parameter optimisation, design-of-experiments
\end{IEEEkeywords}}

\maketitle

\IEEEdisplaynotcompsoctitleabstractindextext

%
\IEEEpeerreviewmaketitle

\section{Introduction}
%
%

%
%
%
%
\IEEEPARstart{N}{owadays} not only engineers and computer scientists use various simulation codes and
data pre- and post-processing tools for almost every task during product development or experimental design.
In the financial sector automated data analysis and predictions from numerical models are common.
Integrating consecutive tasks from distributed systems in workflows frees the user from the need
to transfer data and to invoke applications in due time.

During the last decade middleware such as Globus Toolkit \cite{globus1,globus2,globus3}, 
Unicore \cite{unicore1,unicore2}, and gLite \cite{glite1,glite2} was developed in grid projects. 
These tools mainly focus on service-oriented grid infrastructures with 
seamless and secure access to data sources for users without experience in grid computing.
The middlewares expose applications via web services \cite{webservices} to workflow managers like Taverna workbench \cite{taverna1,taverna2}
enabling users to design and execute complex workflows in graphical interfaces and monitor
the operations remotely. Due to the inherent complexity of abstract database interfaces and
multiple middleware layers specialists are needed to install, configure, and test these tools.
The specialist must be supported by the local administrators of the worker nodes as 
various tasks require administrative rights. Users without programming experience cannot 
deploy data streams in graphical interfaces unless the access to data and applications is encapsulated 
by particular service wrappers. In our experience with scenarios like this developers spent most 
of the time setting up and debugging the diverse tools. The construction of workflows via graphical 
user interfaces, however, is not flexible or efficient enough to save a reasonable amount of time.

As a typical example from mechanical engineering we have chosen numerical simulations where 
input decks containing finite element models have to pass through several workflow stages. 
At each stage an application processes the resulting data from a previous stage to become the 
input data of the subsequent one. In real-world scenarios we face many different applications 
and various restrictions, e.g. a CAD program to create a modified model only runs on a 
dedicated Windows system, while the meshing generator uses an old but reliable Unix system, 
and the finite element simulation code runs on a Linux cluster without sufficient licences for 
all nodes being available. To expose a single service it would not make sense to install Globus Toolkit, 
Unicore, or even gLite on all these nodes. But how can a bunch of legacy software in a highly 
heterogeneous and distributed environment be orchestrated without having administrative rights for the machines?


\section{The Engine}

\begin{figure}
  \begin{center}
    \includegraphics[angle=0,width=0.48\textwidth]{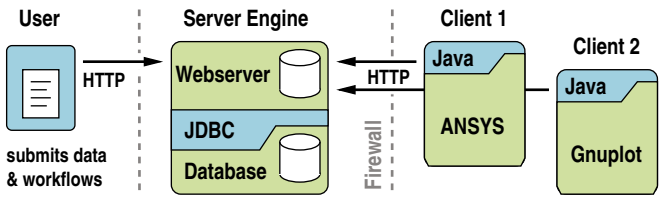}
  \end{center}
  \caption{\label{fig:scenario} The framework in a simple scenario: the server engine 
utilises a database for workflow descriptions and a web server for the communication 
with the clients. Applications are wrapped by lightweight Java clients periodically 
querying the server engine for new tasks.}
\end{figure}

In the framework we cover below, we have inverted the standard scenario of grid computing 
where users act as clients contacting services on distributed hosts. Instead of wrapping 
services on every host by complex middleware for accounting, workflow management, 
and job control we developed a lightweight grid client that can be deployed without installation. 
This client program turns applications on different hosts into remote controllable services. 
For each application a generic wrapper is extended to deal with the application-specific parameters 
and to create and start suitable batch scripts. New wrappers can be added to the system running
without restart. At the central host a servlet container, 
i.e. Apache Tomcat, processes the communication of all clients \cite{servlet,tomcat}. 
A relational database maintains workflow descriptions and parameters for the applications \cite{sql,jdbc}, see Figure~\ref{fig:scenario}.

\subsection{Distributed workflows}

In an exemplary workflow the client downloads data and processes it, i.e. it starts an application 
editing the data. The client monitors the ongoing process and uploads the results after the 
application has finished. Further processing of the data resulting can be delegated to applications 
on other clients. The clients communicate with a servlet container on the central host accessing 
an URL with their logical client name as parameter, e.g.:
\vspace{0.1cm}
\\
\begin{footnotesize}
\begin{tt}
http://server:8180/engine/Tasks?Client=node-01\vspace{0.2cm}
\end{tt}
\end{footnotesize}
\\
The underlying servlet \ttt{Tasks} queries the server database for open tasks 
for this client \ttt{node-01} and generates an XML web page \cite{xml} with the workflow description 
of the tasks. In our simulation example the requesting client receives the detailed
description below as response:

\lstset{basicstyle=\ttfamily\fontsize{6.6}{6.6}\selectfont}
\lstset{language=sh}
\begin{lstlisting} 
<Task ID="1" ClientDir="ansys_001">
 <Job No="1" Type="Download">
  <Parameter ServerDir="models" File="inputdeck_001.dat"/>
 </Job>

 <Job No="2" Type="JobAnsys" Timeout="1000">
  <Parameter Input="inputdeck_001.dat"/>
 </Job>

 <Job No="3" Type="JobParseAnsysEigenfreq" Timeout="1000">
  <Parameter Modefile="eigenmode.asc" Freqfile="eigenfreq.asc"/>
 </Job>

 <Job No="4" Type="Upload">
  <Parameter ServerDir="results/sim_001" File="eigenmode.asc"/>
 </Job>

 <Job No="5" Type="Upload">
  <Parameter ServerDir="results/sim_001" File="eigenfreq.asc"/>
 </Job>
</Task>
\end{lstlisting} 
The engine stores parameters for each type of job in a separate table, 
in our example \ttt{Download}, \ttt{Upload}, \ttt{JobAnsys}, and \ttt{JobParseAnsysEigenfreq}. 
Client applications can be supplied with an arbitrary number of parameters. 
For each parameter a column in the corresponding job table for this type of application has to be created. 
The table columns are the attributes of the embedded \ttt{Parameter} elements.
The database table \ttt{Download} contains paths to the input decks for the Ansys simulations 
in the server file system:

\lstset{basicstyle=\ttfamily\fontsize{6.2}{6.2}\selectfont}
\begin{lstlisting} 
select * from Download;
+-----+--------+-------------------+--------+----------+---------+
| Job | Server |                   |        |          |         |
| ID  | Dir    | File              | Length | MD5Sum   | LastMod.|
+-----+--------+-------------------+--------+----------+---------+
|  11 | models | inputdeck_001.dat |  20592 | fd13fb.. | 2009-.. |
|  12 | models | inputdeck_002.dat |  20592 | dbf2e9.. | 2009-.. |
+-----+--------+-------------------+--------+----------+---------+
\end{lstlisting} 
If available, the client uses the file length, the last-modified time stamp, and the MD5Sum \cite{md5sum} 
to verify the integrity of downloaded files. Repeated downloads of identical files can be avoided by 
comparing them to a local file cache. Table \ttt{JobAnsys} contains the parameters for a wrapper class 
with the identical name starting an Ansys simulation. The job wrappers will be discussed later in 
section \ref{jobwrappers}. Here the input deck filename is the only parameter to the Ansys job wrapper:

\lstset{basicstyle=\ttfamily\fontsize{6.2}{6.2}\selectfont}
\begin{lstlisting} 
select * from JobAnsys;
+-------+-------------------+---------+
| JobID | Input             | Timeout |
+-------+-------------------+---------+
|    11 | inputdeck_001.dat |    1000 | 
|    12 | inputdeck_002.dat |    1000 | 
+-------+-------------------+---------+
\end{lstlisting} 
The generic parameter \ttt{Timeout} is available for all kind of jobs, it is particularly useful 
for tests during the construction of new workflows. Once set, running jobs are automatically 
aborted after a previously defined period, here after 1000 seconds. The next job \ttt{JobParseAnsysEigenfreq} 
parses the Ansys output and creates two files \ttt{Freqfile} and \ttt{Modefile} for natural frequencies 
and mode shapes. The Ansys output filename is not used as a third parameter because the default name is 
already hardcoded in the parser script:

\lstset{basicstyle=\ttfamily\fontsize{6.2}{6.2}\selectfont}
\begin{lstlisting} 
select * from JobParseAnsysEigenfreq;
+-------+---------------+---------------+---------+
| JobID | Modefile      | Freqfile      | Timeout |
+-------+---------------+---------------+---------+
|    11 | eigenmode.asc | eigenfreq.asc |     120 | 
|    12 | eigenmode.asc | eigenfreq.asc |     120 | 
+-------+---------------+---------------+---------+
\end{lstlisting} 
Finally, the table \ttt{Upload} contains the names of the two result files. 
The files are created on the client by the parser script, thus their length and 
MD5sum are unknown before. They are uploaded to the server, column \ttt{ServerDir} creates separate result 
directories in the server file system:

\lstset{basicstyle=\ttfamily\fontsize{6.2}{6.2}\selectfont}
\begin{lstlisting} 
select * from Upload;
+-----+-----------------+---------------+--------+-----+------+
| Job |                 |               |        | MD5 | Last |
| ID  | ServerDir       | File          | Length | Sum | Mod. |
+-----+-----------------+---------------+--------+-----+------+
|  11 | results/sim_001 | eigenmode.asc |        |     |      |
|  12 | results/sim_001 | eigenfreq.asc |        |     |      |
|  13 | results/sim_002 | eigenmode.asc |        |     |      |
|  14 | results/sim_002 | eigenfreq.asc |        |     |      |
+-----+-----------------+---------------+--------+-----+------+
\end{lstlisting} 
In the previous tables each line describes a single job for a client. 
One ore more jobs can be combined to tasks with unique task IDs.
Tasks are the basic workflow elements of our engine. 
Clients either being on hold or active on the jobs of a single task. 
The next table shows an excerpt of the \ttt{Tasks}' table structure. 
The column \ttt{Job} specifies the names of the tables containing the 
job parameters while column \ttt{JobID} indicates the corresponding line 
numbers. The table contains two similar tasks, each with five jobs for 
clients from the \ttt{cluster} group. The first task (\ttt{TaskID=1}) 
consists of a download job, an Ansys simulation, a call of the script 
to parse the result data, and two upload jobs:

\lstset{basicstyle=\ttfamily\fontsize{6.2}{6.2}\selectfont}
\begin{lstlisting} 
select TaskID, Job, JobID, Client, ClientGroup, Status from Tasks;
+------+------------------------+-----+-----+---------+---------+
| Task |                        | Job | Cli | Client  |         |
| ID   | Job                    | ID  | ent | Group   | Status  |
+------+------------------------+-----+-----+---------+---------+
|    1 | Download               |  11 |     | cluster | waiting |
|    1 | JobAnsys               |  11 |     | cluster | waiting |
|    1 | JobParseAnsysEigenfreq |  11 |     | cluster | waiting |
|    1 | Upload                 |  11 |     | cluster | waiting |
|    1 | Upload                 |  12 |     | cluster | waiting |
|    2 | Download               |  12 |     | cluster | waiting |
|    2 | JobAnsys               |  12 |     | cluster | waiting |
|    2 | JobParseAnsysEigenfreq |  12 |     | cluster | waiting |
|    2 | Upload                 |  13 |     | cluster | waiting |
|    2 | Upload                 |  14 |     | cluster | waiting |
+------+------------------------+-----+-----+---------+---------+
\end{lstlisting} 
A task can be assigned to a specific client or to a client group, here \ttt{cluster}. 
In the mapping table \ttt{ClientGroups} each client can be associated with several groups, 
here some nodes of the cluster group are shown:

\lstset{basicstyle=\ttfamily\fontsize{6.2}{6.2}\selectfont}
\begin{lstlisting} 
select * from ClientGroups;
+-------------+---------+
| ClientGroup | Client  |
+-------------+---------+
| cluster     | node-01 | 
| cluster     | node-02 | 
| cluster     | node-03 | 
| cluster     | node-04 | 
| cluster     |   ...   |
+-------------+---------+
\end{lstlisting} 
Tasks are assigned to the clients sorted by lowest available ID 
and status \ttt{waiting}. To block the task for other clients 
the \ttt{Tasks} servlet changes the status to \ttt{active} with 
the following SQL command \cite{sql,jdbc}:

\lstset{basicstyle=\ttfamily\fontsize{7.5}{7.5}\selectfont}
\lstset{language=SQL}
\begin{lstlisting} 
UPDATE Tasks SET Status='active', Client='$client' 
 WHERE TaskID='$taskID';
\end{lstlisting} 
In SQL statements we use '\$'-notation to indicate the values of variables. 
The servlet assembles the SQL statements in string buffers replacing variables by their 
values, here it substitutes the logical name of the requesting client and the recent
task ID. Further SQL statements process all jobs of this task by subsequent queries 
to the tables \ttt{Download}, \ttt{JobAnsys}, \ttt{JobParseAnsysEigenfreq} etc.,
i.e. they select the lines from the column \ttt{JobID} and generate the XML document with 
workflow instructions for the client.

When a client is initially started, the server's IP address, a unique logical name for the client (e.g. \ttt{node-01}),
a local directory (e.g. \ttt{/tmp/node-01/}), and optionally a client group can be specified.  
Downloads, uploads, and all applications only act on the data in this client directory. 
The client in our example creates a working directory \ttt{ClientDir="ansys$\_$001"} for 
the Ansys task within \ttt{/tmp/node-01/}. The optional parameter \ttt{ServerDir} is used to 
relocate files in the directory structure during the up- and downloads, e.g. the input decks 
come from a special model library directory on the server. The Ansys results from different 
clients are also collected on the server in a joint directory \ttt{results/}. 
Clients and server automatically create all directories from the workflow descriptions.

Several clients with different client directories can run on the same machine without interfering.
Independent clients with distinct logical names provide a convenient way to exploit the computing power 
of multicore and multiprocessor systems.

In our example the \ttt{models/} directory on the server comprises a set of input decks. 
The file system is exposed to the clients by the web server via HTTP. 
Clients can download files requesting an URL consisting of the server's IP address, 
the web server subdirectory, and the name of the file. The path information is URL-encoded 
to deal with potential whitespaces in directory or file names. 
For file uploads the clients send HTTP-POST forms and SOAP-with-attachment requests \cite{soap} to the server. 
On the server side incoming multipart/form-data is handled by a particular upload servlet and the web 
service extension Axis2 \cite{axis}. These services directly write uploaded files to the file system of the 
web server making the data immediately available for other clients.

The input decks in the  server's \ttt{models/} directory are part of different Ansys simulation runs. 
The first requesting client downloads \ttt{inputdeck$\_$001.dat} to its local working directory 
\ttt{/tmp/node-01/ansys$\_$001/}, performs an Ansys simulation, and starts a script 
that parses the results to the files \ttt{eigenmode.asc} and \ttt{eigenfreq.asc} in the same directory. 
Table \ttt{TasksWorkingDir} allows to allocate each task to a different working directory on the clients:

\lstset{basicstyle=\ttfamily\fontsize{6.2}{6.2}\selectfont}
\lstset{language=sh}
\begin{lstlisting} 
select * from TasksWorkingDir;
+--------+-----------+-------------+
| TaskID | Dir       | EraseOnExit |
+--------+-----------+-------------+
|      1 | ansys_001 | false       |
|      2 | ansys_002 | false       |
+--------+-----------+-------------+
\end{lstlisting} 
Once the task is finished, the flag \ttt{EraseOnExit} automatically deletes the subdirectory. 
Subsequent tasks can operate on the result data cached in the working directory, if set to false.

\subsection{Job wrappers for local applications \label{jobwrappers}}

Basic jobs for downloading and uploading data are integrants of the grid client.
A workflow engine must be capable of wrapping further applications running on the clients. 
In our framework a generic class \ttt{Job} is extended to wrap new applications. 
The class name of the extended job wrapper corresponds to the job name in the \ttt{Tasks} table. 
The grid client automatically starts the job wrapper and supplies it with the job parameters of the XML description. 
The extended wrapper class employs these parameters and assembles a valid script calling 
the application. The script is temporarily generated in an internal string buffer. 
A method of the generic job class replaces all aliases (see below) and writes it to the 
working directory of the client. The grid client uses Java Native Interface (JNI) to start 
the script and monitor the activity of the application \cite{JNI}. In our example, 
the string buffer created by \ttt{JobAnsys} looks like this:

\lstset{basicstyle=\ttfamily\fontsize{8}{8}\selectfont}
\lstset{language=sh}
\begin{lstlisting} 
#!/bin/bash
cd /tmp/node-01/ansys_001/
export LM_LICENSE_FILE=ANSYS_LICENSE
ANSYS -b -i inputdeck_001.dat
\end{lstlisting} 
The shell header and the change-the-directory command in line 1 and 2 are 
selected by the generic job class according to the operating system of the client. 
Further lines are created by the extended wrapper class, \ttt{ANSYS$\_$LICENSE} here 
defines a so-called alias that can be understood as an internal environment variable. 
Before the script is written to the working directory the generic job wrapper 
contacts an \ttt{Aliases} servlet on the server and requests a list of known aliases for 
the given client. It then replaces all aliases by machine-specific parameters.
The highst priority have aliases directly attached to a 
client's name, e.g. \ttt{JAVA} on client \ttt{node-01} is mapped 
to \ttt{/opt/jdk1.6.0$\_$12/bin/java}. Then the aliases for entire groups 
of clients are exchanged. Finally, global aliases not assigned to a 
client or client group (e.g. \ttt{JAVA} is mapped to \ttt{java}) 
are applied:

\lstset{basicstyle=\ttfamily\fontsize{6.2}{6.2}\selectfont}
\begin{lstlisting} 
select * from Aliases;
+---------------+-----------------------------+---------+---------+
|               |                             | Client  |         |
| ServiceName   | Service                     | Group   | Client  |
+---------------+-----------------------------+---------+---------+
| ANSYS         | /opt/ansys/v11/bin/ansys11  | cluster |         | 
| ANSYS_LICENSE | /opt/ansys/shrd/license.lic | cluster |         | 
| JAVA          | /usr/lib/jdk1.6.0/bin/java  | linux   |         | 
| JAVA          | /opt/jre-6-solaris/bin/java | solaris |         | 
| JAVA          | /opt/jdk1.6.0_12/bin/java   |         | node-01 | 
| JAVA          | /opt/jdk1.6.0_12/bin/java   |         | server  | 
| JAVA          | java                        |         |         | 
| GNUPLOT       | /usr/bin/gnuplot            | linux   |         | 
| GNUPLOT       | /opt/gnu/bin/gnuplot        | solaris |         | 
+---------------+-----------------------------+---------+---------+
\end{lstlisting} 
The next listing shows the extended job wrapper class for Ansys. The grid client creates a job class instance, 
assigns it with the logical client name, the working directory \ttt{dir}, and Ansys-specific parameters, 
and starts the \ttt{run()} method in a separate thread. The call of \ttt{replaceAliases()} exchanges the aliases 
\ttt{ANSYS} and \ttt{ANSYS$\_$LICENSE} in the generic wrapper script by machine-specific values from 
the \ttt{Aliases} table. Based on the client's name the \ttt{Aliases} servlet returns a list of all known 
aliases as key-value-pairs. Function \ttt{createScript()} writes the completed script to a file 
\ttt{script$\_$ansys} in the client working directory. 
The thread waits at \ttt{executeScript()} until the Ansys simulation has finished:

\lstset{basicstyle=\ttfamily\fontsize{6.8}{6.8}\selectfont}
\lstset{language=Java}
\begin{lstlisting} 
public class JobAnsys extends Job
{
  String inputdeck;   // Parameter for Ansys

  public void setParameters(Element parameter)
  {
    inputdeck = getAttribute(parameter, "Input");
  }

  public void run()
  {
    log("Starting JobAnsys for " + dir + "/" + inputdeck);

    StringBuffer t = new StringBuffer();
    t.append("export LM_LICENSE_FILE=ANSYS_LICENSE" + "\n");
    t.append("ANSYS  -b  -i " + inputdeck           + "\n");

    Script s = new Script(dir, "script_ansys");
    s.createScript( getShellHeader() 
                  + replaceAliases(t.toString()) );
    s.executeScript();

    if (! (new File(dir, "mode4.png")).exists() ) 
    {
      log("- Job failed -");
      setStatus(Job.JOB_ANSYS_FAILED);
    }

    flag_done = true;   // Signal to the main thread
  }
}
\end{lstlisting} 
At the same time the main thread monitors process load and disk space of the client in a loop. 
A boolean \ttt{flag$\_$done} notifies the main thread that the Ansys job is finished. 
In our example the Ansys input decks contain additional instructions to create images 
of the simulated finite element models. The wrapper uses image \ttt{mode4.png} to 
ascertain whether the simulation was successfully finished and data for further processing 
is available.

The job wrapper automatically redirects standard output and error streams to a 
log file in the working directory. With some applications running separately 
in a background thread, e.g. PamCrash impact simulations, only the content 
of the log file can be analysed to monitor the progress and determine when 
the job will be finished. The job wrapper passes the final status of the job 
execution by an integer number to the grid client, indicating whether the 
simulation run has been successfully or not. After all jobs of a task are 
completed, the grid client contacts the servlet \ttt{TaskCompleted} on the 
central host and sends task ID, status information, and additional information 
(total runtime of the task, the recent process load and free disk space) 
to the server. 

The status information allows the engine to decide whether to redistribute the same task, 
e.g. to a different client, or to tick it off and unlock further tasks depending on it 
(see section \ref{task_chains}, Task chains). If a client issues a request for new tasks 
without having completed the last active task by calling \ttt{TaskCompleted}, e.g. 
because the process was interrupted or the client was switched off, the status of 
the former task is again set back to \ttt{waiting}. The \ttt{Tasks} servlet 
implements this by an SQL update:

\lstset{basicstyle=\ttfamily\fontsize{7.5}{7.5}\selectfont}
\lstset{language=SQL}
\begin{lstlisting} 
UPDATE Tasks SET Status='waiting' 
 WHERE Client='$client' AND Status='active';
\end{lstlisting} 
The interrupted client then resumes its work and restarts the task.

\subsection{Parameter template files \label{template_parameter_files}}

Sometimes command line arguments are not sufficient and special parameter files are
needed to control an application. Before the application is started parameter files 
can be downloaded from the central server or a particular job can directly create them 
on the client. For the download case we have developed a generic job that substitutes 
embedded tags in a parameter template file by numerical values from a list. 
Parameter template files are useful in design-of-experiment studies, numerical optimisations, 
or analysis of the robustness of a solution, where many different sets of parameters have to be tested.

Therefore the pre-defined template file is deployed on the server and the job \ttt{ReplaceTags} 
generates a valid parameter file for a simulation run from the downloaded template. 
This job parses the template and looks for placeholders in XML-element form. The names of the 
elements can be arbitrarily chosen to avoid collisions with keywords of the parameter file. 
The mandatory attribute \ttt{ID} serves as a key to replace the entire tag by a numerical value 
from the parameter list of the \ttt{ReplaceTag} job. To handle fixed-format files with strict column widths
an optional attribute \ttt{Len} specifies the number of characters, 
the numerical values have to be accordingly rounded down or extended with whitespaces.
The XML tags can be extended by further attributes, i.e. we have determined minimum and maximum bounds 
(or mean value and deviation) for each parameter. This enables \ttt{ReplaceTag} to substitute 
the placeholders by random values from the defined range (or values from a normal distribution around the mean value). 
This feature allows the user to select and alter arbitrary parameters in design-of-experiment studies on the fly 
without any changes to the workflow.

The next listing provides an excerpt of an input deck template for Ansys \cite{ansys} that we use 
in tests with finite element simulations. The first line defines a square solid plate with 
edge length $a$, the next three lines construct solid cylinders at fixed positions inside the plate. 
In the input deck we substitute the radii of the cylinders by tags with identifiers \ttt{r1}, \ttt{r2}, 
and \ttt{r3}. The last line drills holes into the structure subtracting the cylinder shapes from the plate (cf. Fig. \ref{fig:plate_fem}):

\lstset{basicstyle=\ttfamily\fontsize{6.9}{6.9}\selectfont}
\lstset{language=sh}
\begin{lstlisting} 
RECTNG,0,a,0,a
CYL4,32e-3,32e-3,<TAG ID="r1" Min="1e-3" Max="7e-3" Len="5"/>
CYL4,28e-3, 9e-3,<TAG ID="r2" Min="1e-3" Max="7e-3" Len="5"/>
CYL4,10e-3,30e-3,<TAG ID="r3" Min="1e-3" Max="7e-3" Len="5"/>
ASBA,1,ALL
\end{lstlisting} 
With this template we explore the behaviour of the plate for holes with random diameters in the 
feasible range from \ttt{Min}=1 to \ttt{Max}=7 mm (see section~\ref{applications:doe} for more details).
In a second study the radii are additionally subject to an optimisation procedure that attempts to 
match the simulated natural frequencies of the plate with a given spectrum (see section \ref{parameter_optimisation}).
To keep track of the parameters actually used in the studies \ttt{ReplaceTag} creates a log file 
for each simulation run. The log file and the resulting frequencies from the Ansys simulation
are uploaded to a result directory. A client on the server initiates a job to add the 
new information to a database table maintaining all parameter and result values of the studies.

\subsection{Joint services across mixed operating systems}

The workflow engine can merge identical applications and functionalities from loosely 
coupled heterogeneous clients as abstract services. The engine provides a transparent 
access of services as resources in the grid-independent of the underlying client's operating systems. 
In general the aliases mechanism is not sufficient as not only the parameters but the scripts 
themselves differ on different platforms. To simplify the maintenance of the system we have put 
all different scripts for the utilised operating systems in the same job wrapper class so that 
the wrapper can switch between them. The workflow engine maintains a mapping table, specifying an 
operating system for each client name.

The complete Ansys wrapper contains two script templates for Windows and Linux.
These templates only differ in an \ttt{export} versus a \ttt{set} command while utilising 
identical aliases for the Ansys binary and the license server. In the next step we have introduced
a new client group \ttt{ansys} selecting all machines able to run Ansys. Tasks assigned 
to this group are automatically distributed to all requesting grid clients independent of their operating system.

Further client dependent parameters of services can again be handled with aliases, an alias \ttt{JAVA$\_$OPTIONS} e.g. 
specifies the maximum available heap size for a Java process to \ttt{-Xmx4000M} on a particular machine. 
If we subsume all machines with a large amount of memory in a new group \ttt{ansys-large}, while they are still 
members of the normal \ttt{ansys} group, they participate in every Ansys simulation 
whereas extensive simulation runs with bulky input decks can be exclusively allocated to the new client group.

The combination of client groups, aliases lists, and job wrappers with operating system switches allows 
the use of abstract services in the workflow description table \ttt{Tasks} whereas the actual implementations 
of services differ on the clients.

\subsection{Task chains \label{task_chains}}

Complex workflows can be constructed with task chains realised by a 
special column \ttt{DependsOnTask} in the tasks table. The column maintains the task ID of a 
predecessor task. In the following example tasks 1 and 3 can be immediately executed by clients
from the cluster group while tasks 2 and 4 for a client running on the server depend on
the results of the preceding Ansys simulations:

\lstset{basicstyle=\ttfamily\fontsize{6.2}{6.2}\selectfont}
\begin{lstlisting} 
select TaskID, Job, JobID, DependsOnTask, Client, ClientGroup 
  from Tasks;
+------+------------------------+-----+-------+--------+---------+
| Task |                        | Job | Dep's |        | Client  |
| ID   | Job                    | ID  | OnTask| Client | Group   |
+------+------------------------+-----+-------+--------+---------+
|    1 | Download               |  40 |       |        | cluster |
|    1 | JobAnsys               |  21 |       |        | cluster |
|    1 | JobParseAnsysEigenfreq |  21 |       |        | cluster |
|    1 | Upload                 |  80 |       |        | cluster |
|    1 | Upload                 |  81 |       |        | cluster |
|    2 | Download               |  41 |     1 | server |         |
|    2 | InsertIntoDatabase     |  11 |     1 | server |         |
|    3 | Download               |  42 |       |        | cluster |
|    3 | JobAnsys               |  22 |       |        | cluster |
|    3 | JobParseAnsysEigenfreq |  22 |       |        | cluster |
|    3 | Upload                 |  82 |       |        | cluster |
|    3 | Upload                 |  83 |       |        | cluster |
|    4 | Download               |  43 |     3 | server |         |
|    4 | InsertIntoDatabase     |  12 |     3 | server |         |
+------+------------------------+-----+-------+--------+---------+
\end{lstlisting} 
One action of the servlet \ttt{TaskCompleted} is to set the status of a completed task to \ttt{done}
and to remove the respective task ID from all further entries in the \ttt{DependsOnTask} column: 

\lstset{basicstyle=\ttfamily\fontsize{6.3}{6.3}\selectfont}
\lstset{language=SQL}
\begin{lstlisting} 
UPDATE Tasks SET Status='done'      WHERE        TaskID='$taskID';
UPDATE Tasks SET DependsOnTask=NULL WHERE DependsOnTask='$taskID';
\end{lstlisting} 
This procedure is repeated for each task completed and activates all tasks depending on the one just solved. 
Servlet \ttt{Tasks} then assigns the first available task to a requesting client by the following scheme:

\begin{lstlisting} 
CREATE TEMPORARY TABLE TasksTemp  SELECT * from Tasks    
 WHERE Status='waiting' AND DependsOnTask IS NULL ORDER BY TaskID; 
CREATE TEMPORARY TABLE TasksTemp2 SELECT * from TasksTemp 
 WHERE Client IS NULL ORDER BY TaskID;

SELECT * FROM TasksTemp  WHERE Client='$client';
SELECT * FROM TasksTemp2 T, ClientGroups C 
 WHERE T.ClientGroup=C.ClientGroup AND C.Client='$client';
SELECT * FROM TasksTemp2 WHERE ClientGroup IS NULL;
\end{lstlisting} 
The first two statements create temporary tables only for tasks whose 
status is \ttt{waiting} and if \ttt{DependsOnTask} is not set.
This prevents tasks reliant on predecessors tasks from being distributed and speeds up all subsequent selections. 
The first \ttt{select} statement looks for tasks directly assigned to the logical name of a requesting client. 
The second statement selects tasks for all groups the requesting client is member of. 
The last statement only acts on tasks neither assigned to a particular client nor to a client group. 
The tasks are ordered by their IDs, the servlet returns the task with the lowest ID to the client. 
This sequence of select queries prioritises tasks bound to a specific client against tasks for an entire client group. 
It prioritises workflows waiting for a special service available only on a single client thus avoiding 
bottlenecks as such a workflow does not have to wait until all further tasks for the client group are completed, 
e.g. typically extensive simulation runs intended for a cluster group.

The initial status of the task chains is \ttt{passive} to prevent clients from working on incomplete chains,
where premature calls to \ttt{TaskCompleted} cannot free successor tasks that have not yet been inserted. 
After the last task of a chain has been inserted the chain is enabled by switching the status to \ttt{waiting}. 
The SQL commands are executed during every client request. To ensure database performance for huge numbers of 
tasks we have created indexes on the columns \ttt{TaskID}, \ttt{DependsOnTask}, \ttt{Client}, 
\ttt{ClientGroup} and \ttt{Status}.

\subsection{Client monitoring}

As process load and available disc space cannot directly be monitored by a Java virtual machine,
we have integrated a monitor to apply native functions such as \ttt{uptime} and \ttt{df} on Linux 
based systems. The clients automatically supply the server with this information during queries 
for new tasks:
\vspace{0.1cm}
\\
\begin{scriptsize}
\begin{tt}
http://.../Tasks?Client=node-01\&Load=1.54\&Disk=73360476\vspace{0.2cm}
\end{tt}
\end{scriptsize}
\\
If no job is available the client sleeps for a certain interval before repeating its request.
Sleep time intervals can be independently specified in the table \ttt{ClientsAvailable}. 
The load and disk space information are also monitored here:

\lstset{basicstyle=\ttfamily\fontsize{6.2}{6.2}\selectfont}
\lstset{language=sh}
\begin{lstlisting} 
select Client, IP, TaskID, LastRequest, Sleeptime, 
       Load, Diskspace from ClientsAvailable;
+---------+----+------+----------------+-------+------+----------+
|         |    | Task |                | Sleep |      | Disk     |
| Client  | IP | ID   | LastRequest    | time  | Load | space    |
+---------+----+------+----------------+-------------------------+
| node-01 | .. |   -1 | 06-23 11:23:08 |   300 | 1.54 | 73360476 |
| node-02 | .. |   -1 | 06-23 11:25:01 |   300 | 1.79 | 71360000 |
| node-03 | .. |   -1 | 06-23 11:25:21 |   300 | 0.69 | 69359261 |
| node-04 | .. |   -1 | 06-23 11:25:23 |   300 | 1.72 | 71559921 |
| server  | .. |   -1 | 06-23 11:25:46 |    30 | 0.32 | 86659732 |
+---------+----+------+----------------+-------+------+----------+
\end{lstlisting} 
In addition, each client's request automatically updates the column \ttt{LastRequest}.
The columns \ttt{LastRequest} and \ttt{Sleeptime} allow to estimate when the client will again contact the server. 
Clients exceeding this time interval may be down or cannot connect with the server.

With Secure Shell (SSH) \cite{ssh} active clients can start further clients on remote machines. 
Via SSH access the client software itself can also be distributed, e.g. to all nodes of a cluster. 
Table \ttt{Clients} maintains the information needed: user@client, password or key file,
location of the installation directory the user is allowed to write in, 
and the group of clients able to access the remote system. 
We integrated a job \ttt{StartGridClient} that utilises the pure Java SSH implementation JSch \cite{jsch} 
to copy the local installation to remote machines and execute it there. 
To apply this method to all clients of a cluster, e.g. after a reboot, we can 
insert \ttt{StartGridClient} jobs for all clients currently not available by querying 
the tables \ttt{Clients} and \ttt{ClientsAvailable}. Job submission is discussed later 
in section \ref{parameter_optimisation}. After an initial client has been started manually 
it automatically installs and starts further clients. All clients started like this are active
in the process with the number of active grid clients doubling at each stage. 
We so managed to install and start clients on a cluster with 100 nodes in less than two minutes. 
This mechanism is also especially useful to redistribute the entire client system after major 
modifications.

\subsection{Dynamic class loading}

All job wrappers for client applications are maintained in a special package directory 
on the client. Extensions to or modifications of these wrappers are frequent practise. 
To avoid a complete redistribution of the system after wrapper specific modifications
we have additionally introduced a directory on the web server for compiled job wrapper 
classes. The clients utilise dynamic class loading \cite{rmi} for inherited classes of 
\ttt{Job} except the very basic ones. 

If a class for a specific job name from a workflow description, e.g. \ttt{JobAnsys}, 
cannot be found in the local repository of the client's JAR file the client attempts to download and 
instantiate the wrapper class from the central web server directory. Thus new application wrappers 
and modifications are immediately visible to all clients within the grid system and a redistribution 
is possible without having to restart.

\subsection{Administrative rules}

The assingment of clients to client groups can be dynamically changed in the central database 
according to administrative rules. One rule states that e.g. next week a subset of cluster nodes is 
booked for benchmarks and cannot be accessed during the day. Beside the existing group 
\ttt{cluster} for all nodes, for the subset we have introduced a new group \ttt{reserved}.
We then establish cron jobs \cite{cron1,cron2} on the server to have all reserved clients from the cluster
group removed in the morning and re-assigned in the evening:

\lstset{basicstyle=\ttfamily\fontsize{6.1}{6.1}\selectfont}
\lstset{language=SQL}
\begin{lstlisting} 
# Cron job at 7 am:
# 
CREATE TEMPORARY TABLE ReservedNodes SELECT Client FROM ClientGroups 
 WHERE ClientGroup="reserved";
UPDATE ClientGroups SET ClientGroup="cluster_suspended"
 WHERE ClientGroup="cluster" AND 
       Client IN (SELECT Client FROM ReservedNodes);
UPDATE ClientsAvailable SET SleeptimeMin=3600, SleeptimeMax=3600 
 WHERE Client IN (SELECT Client FROM ReservedNodes);

# Cron job at 7 pm:
# 
UPDATE ClientsAvailable SET SleeptimeMin=60, SleeptimeMax=300 
 WHERE Client IN (SELECT Client FROM ClientGroups 
                   WHERE ClientGroup="cluster_suspended");
UPDATE ClientGroups SET ClientGroup="cluster" 
 WHERE ClientGroup="cluster_suspended";
\end{lstlisting} 
Jobs can now be scheduled to the same cluster as before, at night the cluster is automatically extended 
by the nodes reserved for benchmarks. This set-up is especially useful if runtime and end of a jobs can be 
estimated beforehand as e.g. in the case of many identical simulations such as parameter studies or optimisation runs.

The first script also prolongs the request interval for the reserved nodes up to 1 hour at daytime. 
This further reduces the activity on the benchmark nodes. It is possible to provide different minimum and maximum values 
for the sleep time interval, e.g. 1 to 5 minutes at night. The server randomly picks a value out of 
this interval. By these random time offsets we avoid long term synchronisations, if many grid clients 
simultaneously request a new task.

We applied long request intervals between 120 and 10,000 seconds ($\approx 3$h) in tests with clients 
on 85 different cluster nodes resulting in statistically one node per minute requesting a new job. 
If enough jobs are available an increasing number of nodes gets involved. 
The client system implicitly realises a load balancing with Round-Robin scheduling \cite{roundrobin1,roundrobin2}
and automatically utilises more nodes for long running processes. After approximately 90 minutes 
the entire cluster is involved.

The same technique is applied to distribute tasks to idle cluster nodes. 
With every request the clients update their current load levels in 
the \ttt{ClientsAvailable} table. A cron job periodically selects clients 
with load levels above individual thresholds, removes them from their regular client groups 
and puts them on hold as long as their load level does not change. 
This does not effect tasks allocated to a particular client name, 
these are further executed, even if the load level of a specific client is high.

\section{Applications \label{applications}}

In this section a first implementation of the previously described 
workflow engine is used in design-of-experiment studies, parameter optimisations, 
and robustness validations of finite element structures. We propose 
extensions for workflow submission and stateful services and discuss 
interfaces to integrate algorithms and external libraries to the modular system.

\subsection{Design-of-experiment studies\label{applications:doe}}

The first workflow deals with parameter variations and can be used in design-of-experiment (DOE) 
studies \cite{doe2}. It includes pre- and post-processing jobs to maintain all results in database tables. 
To save simulation time we employ a toy model from mechanical engineering for the tests.
The finite element code Ansys \cite{ansys} is used to simulate the natural frequencies of a square plate containing holes 
of variable diameter. Generally, vibrational analysis is applied in structural investigations to meet the natural 
frequencies of a system with certain requirements. In automotive development the natural frequencies 
of the vehicle body are not allowed to overlap with excitations by aggregates or road-wheel interactions.
In the development of vibration sensors the natural frequencies must be robust against parameter inaccuracies, 
due to e.g. inhomogenities of the material or temperature fluctuations, and the exact shape of the eigenmodes 
is essential to develop reliable electrical transducers.

\begin{figure}
\includegraphics[angle=0,width=0.23\textwidth]{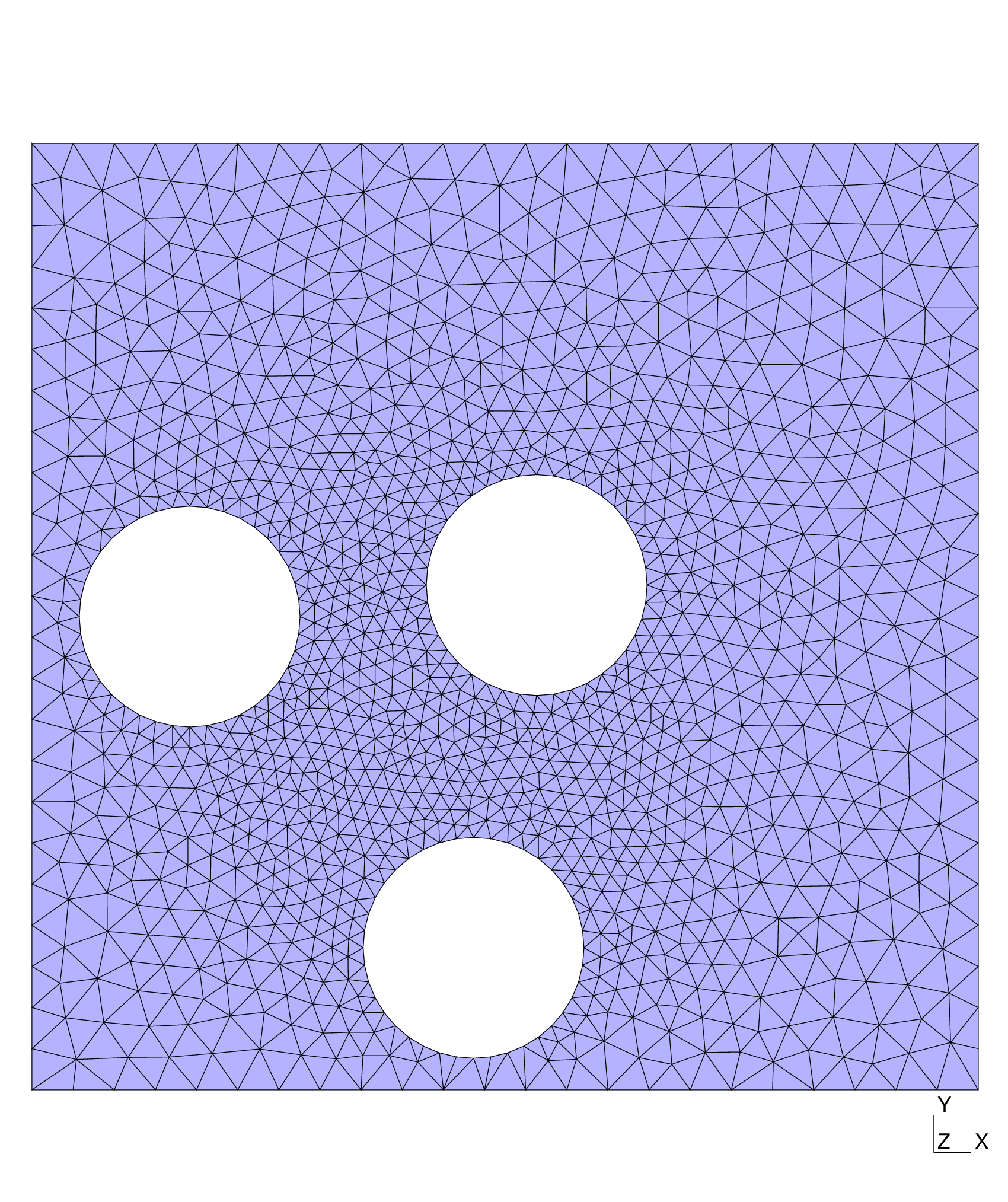} \hspace{0.5cm}
\includegraphics[angle=0,width=0.23\textwidth]{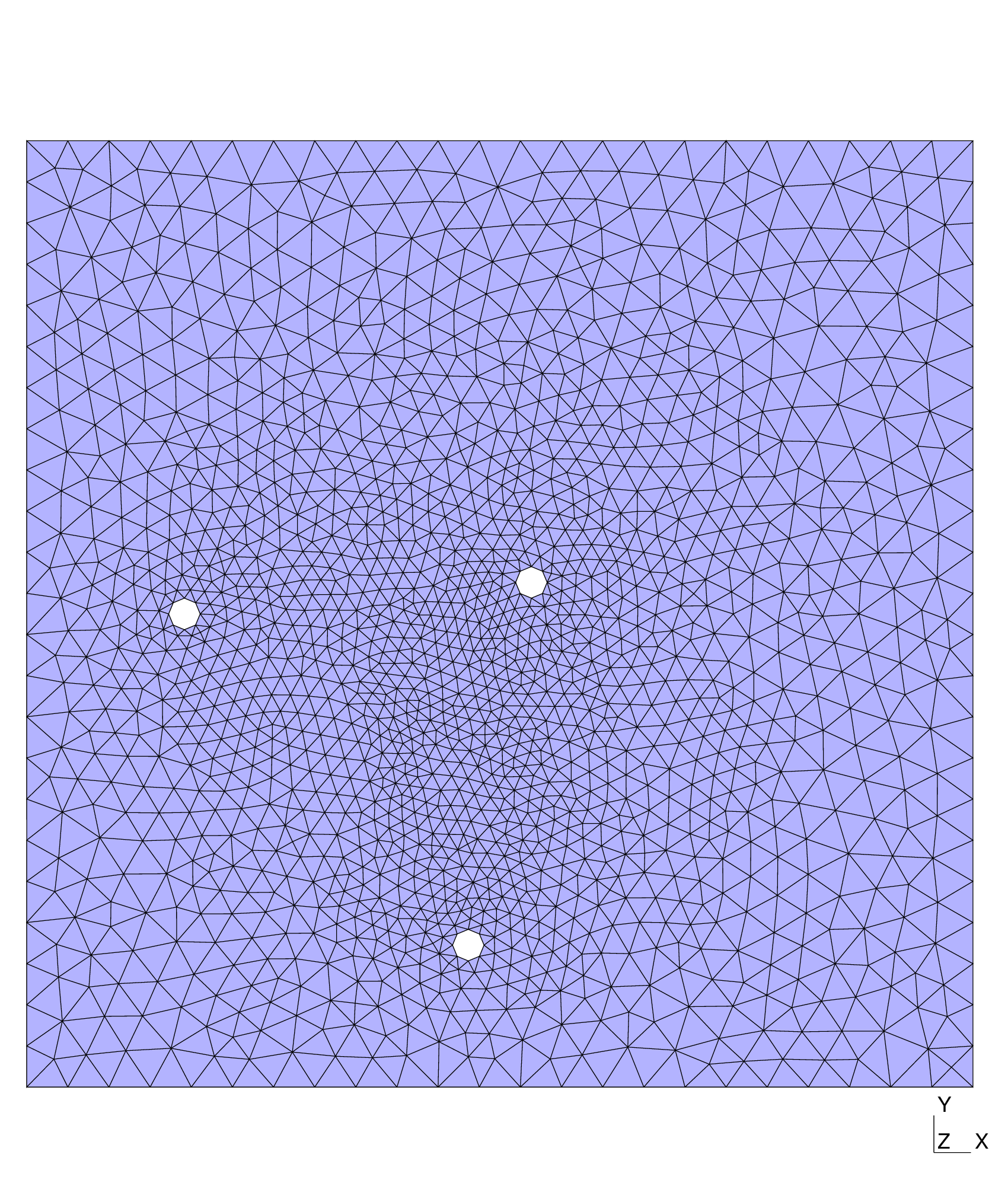}
\caption{\label{fig:plate_fem} Two designs of the 60$\times$60~mm plate as finite element models. 
The radius is limited to a maximum of 7~mm so that the holes will not protude the plate's edges (left), 
a minimum 1 mm was introduced because the meshing generator cannot process vanishing holes (right). 
In experimental design and optimisation studies the hole radii vary independently.}
\end{figure}

\begin{figure}
\vspace*{-0.3cm}
  \hspace*{-0.5cm}	
  \includegraphics[angle=0,width=0.54\textwidth]{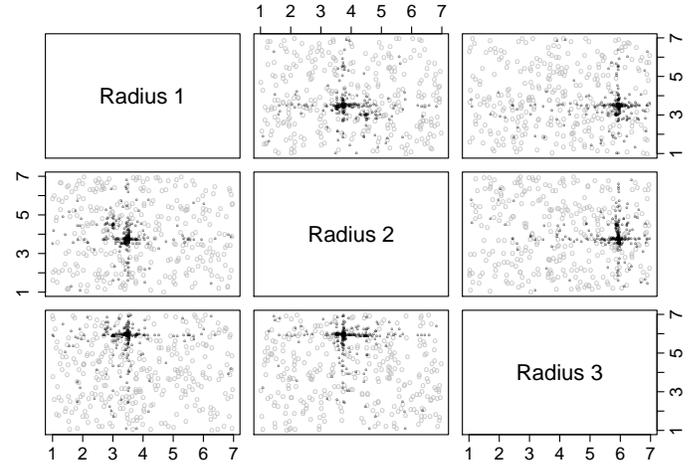}
  \vspace*{-0.8cm}	
  \caption{\label{fig:parameters} Scatterplots of parameters: For the first study we pick 300 sets 
of independent hole radii $(r_1, r_2, r_3)={\rm \bf r}$ from the interval $1\ldots7$~mm, grey dots. 
Figure~\ref{fig:modes} shows the 
resulting natural frequencies of the plate. The results are strongly correlated since each hole radius influences 
all frequencies. Black dots are part of a second study, where an optimisation algorithm fits the radii in order 
to produce a given frequency spectrum, i.e. ${\rm \bf f^*} = (28, 30, 33)$~kHz for the lowest modes. 
A plate with ${\rm \bf r^*} = (3.521, 3.755, 5.937)$~mm meets the requirements. Robustness against 
parameter inaccuracies of this design is checked in further data explorations in 
section~\ref{robust_design}.
}
\end{figure}

In the first study we explore the overall behaviour of the solid plate with three holes at fixed positions 
and hole radii between 1 and 7 mm. Figure~\ref{fig:plate_fem} shows finite element models of the plate with maximum and minimum
hole diameters.
\begin{figure}
\includegraphics[angle=-90,width=0.50\textwidth]{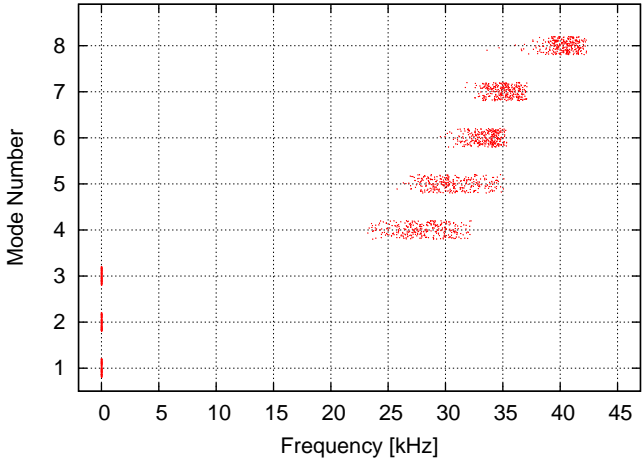}
\caption{\label{fig:eigenmodes_random_doe} Natural frequencies of the plate from Fig.~\ref{fig:plate_fem} 
with holes of varying radii at fixed positions (material density 7.83 g/cm$^3$, Young's modulus 2.1$\cdot 10^{5}$ N/mm$^2$, Poisson's ratio 1/3). 
Ordinate values are extended to enhance visibility. The unsupported plate was simulated in two dimensions, 
the first three modes correspond to rigid body motions with 0~Hz: two translational and one rotational degrees of freedom.}
\end{figure}
We define tags for the radii in the input deck as described in section \ref{template_parameter_files} and create 300 instances of an Ansys workflow. 
A \ttt{ReplaceTag} job exchanges the tags for randomly distributed values from the given interval. 
Figure~\ref{fig:eigenmodes_random_doe} shows the result of 300 simulation runs: the natural frequencies of the lowest 
modes of the vibrating plate. 
Figure~\ref{fig:parameters} depicts the parameters of the hole radii, chosen randomly by 
\ttt{ReplaceTag}, as grey dots. Since these parameter sets cover the entire design space, 
they can be used to construct a Response Surface Model (RSM) of the plate system to predict 
new results without further simulations \cite{rsm}. In section \ref{robust_design} we apply
a similar approach with direct interpolation on the data.
 
Alternative approaches to fill the parameter space are Factorial Designs \cite{factorial_design} 
and Latin Hypercube Designs \cite{doe}, which avoids redundant parameter sets and constructs a RSM 
with similar accuracy in less simulation runs. In such DOEs typically all parameter sets are defined 
before the first simulation is run. As shown in section \ref{template_parameter_files}, the \ttt{ReplaceTag} can 
also handle this DOE with parameter values deriving from a table of the workflow system.

\begin{figure}
  \hspace*{-0.5cm}	
  \includegraphics[angle=0,width=0.54\textwidth]{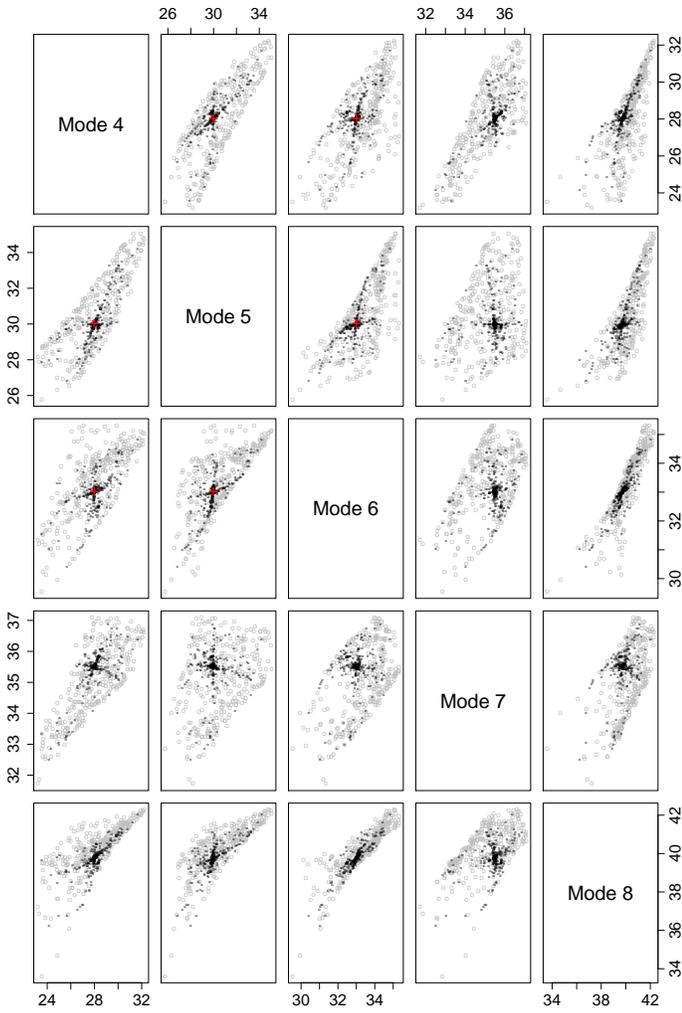}
  \vspace*{-0.8cm}	
  \caption{\label{fig:modes} Scatterplots of the natural frequencies in kHz for all pairs of modes: 
Grey dots indicate the results of the random parameter study to explore 
the solution space (cf. Fig.~\ref{fig:parameters}). The modes are strongly 
correlated as smaller holes increase all frequencies. A mode describes a certain shape of vibration. 
The eigenmode solver sorts all mode shapes by frequency so the numbering of modes with similar frequencies 
interchanges prevalently. These permutations cause broad scatter bands with accumulations on the edges 
(see e.g. mode pair 6 and 7). The black dots belong to the optimisation study, where the first 
three natural modes have to meet the specified frequencies (red crosses).}
\end{figure}

The scatterplots in Figure~\ref{fig:modes} show the resulting frequency combinations for the natural modes of the plate. 
Even though the parameter space is uniformly sampled with random points, the calculated spectrum in the frequency space is limited to 
certain band structures. The transformation from parameters ${\rm \bf r}$ to frequencies ${\rm \bf f}({\rm \bf r})$ involves the full numerics of 
a finite element simulation. In practise this function is not invertible, i.e. no simple rule can be applied to 
find parameter sets for a given frequency set.

\subsection{Parameter optimisation \label{parameter_optimisation}}

We introduce a new package for workflow submission to extend the system by optimisation libraries.
The challenge was to find a special design of the plate, i.e. a certain set of hole radii to produce 
a given eigenfrequency spectrum. Not all combinations of natural frequencies can simultaneously be realised 
as demonstrated in Figure~\ref{fig:modes}. We opt for a feasible set of frequencies 
for the lowest three modes of $f_4^* = 28$, $f_5^* = 30$, and $f_6^* = 33$ kHz exactly. 
Next we look for a parameter set ${\rm \bf r}=(r_1, r_2, r_3)$ to minimise the objective function 
defined as the square distance from the actual frequencies ${\rm \bf f}$ to the given frequency set ${\rm \bf f}^*$:
\begin{equation} \label{objective_function}
  o({\rm \bf r}) =  \left| {\rm \bf f}({\rm \bf r}) - {\rm \bf f}^* \right|^2
\end{equation}
A generic optimisation algorithm proposes an initial parameter set ${\rm \bf r}$ and expects the value of the objective function $o({\rm \bf r})$ back.
Based on the result the algorithm suggests further parameter sets presumably closer to the objective's function minimum.
To integrate optimisation libraries in the workflow system we have developed a package to submit workflows via HTTP 
including job parameters and table descriptions from external applications. The syntax of the workflow 
description is similar to the XML response of the \ttt{Tasks} servlet, but without the concrete numbering for tasks 
and jobs. 

The following script shows an excerpt of a such a workflow description used in DOE studies 
and optimisation runs: a template \ttt{inputdeck.dat} is downloaded from the 
server directory \ttt{models/}, the tags \ttt{r1}, \ttt{r2}, and \ttt{r3} of the 
input deck are then replaced by the numerical values. 
After completion of the input deck Ansys is started. \ttt{JobParseAnsysEigenfreq} parses 
the natural frequencies from Ansys' output. 
An upload job sends the data to the server directory \ttt{results/sim$\_$001/}:

\lstset{basicstyle=\ttfamily\fontsize{6.3}{6.3}\selectfont}
\lstset{language=sh}
\begin{lstlisting} 
<Task ClientDir="ansys_001" startDependsOnClientGroup="cluster"
                            startTaskChain           ="true" 
                            startDependsOnGroupNode  ="true">
 <Job Type="Download">                       
  <Download Dir="models" File="inputdeck.dat"/>
 </Job>                                            
 <Job Type="JobReplaceTag">;
  <Column Name="Input"  Type="VARCHAR" Value="inputdeck.dat"    />
  <Column Name="Output" Type="VARCHAR" Value="inputdeck_mod.dat"/>
  <Column Name="r1"     Type="DOUBLE"  Value="2.83535"          />
  <Column Name="r3"     Type="DOUBLE"  Value="3.64375"          />
  <Column Name="r3"     Type="DOUBLE"  Value="6.21132"          />
 </Job>
</Task>

<Task ClientDir="ansys_001">
 <Job Type="JobAnsys">
  <Column Name="Input" Type="VARCHAR" Value="inputdeck_mod.dat"/>
 </Job>                                            
</Task>

<Task ClientDir="ansys_001">
 <Job Type="JobParseAnsysEigenfreq">
  <Column Name="Freqfile" Type="VARCHAR" Value="eigenfreq.asc"/>
  <Column Name="Modefile" Type="VARCHAR" Value="eigenmode.asc"/>
 </Job>
 <Job Type="Upload">                       
  <Upload Dir="results/sim_001" File="eigenfreq.asc"/>
 </Job>                                            
</Task>

<Task ClientDir="temp" startDependsOnClient="server">
 <Job Type="Download">                       
  <Download Dir="results/sim_001" File="eigenfreq.asc"/>
 </Job>                                            
 <Job Type="JobInsertAnsysResultIntoDatabase">;
  <Column Name="Tablename" Type="VARCHAR" Value="AnsysResults" />
  <Column Name="Datafile"  Type="VARCHAR" Value="eigenfreq.asc"/>
  <Column Name="SimID"     Type="INTEGER" Value="001"          />
 </Job>
</Task>
\end{lstlisting} 
The last task is assigned to a grid client on the server with access to the local database via JDBC \cite{jdbc}, confer Fig.~\ref{fig:scenario}.
Like every other client it first downloads the results to a local working directory \ttt{temp/} before 
a second job inserts them to the table \ttt{AnsysResults}. The directory extension "\ttt{\_001}" is incremented 
in each simulation run, in the result table it acts as a unique ID.

The workflow submission package provides a generic way to assemble the XML description and convey 
it to the server. A workflow submission client is designed to work within the loop of an optimisation algorithm. 
It is triggered by the subroutine that usually evaluates the objective function. During the first call 
JNI Invocation \cite{JNI} initialises a persistent instance of a virtual machine within the C/C++ or 
Fortran subroutine to hold the Java submission client. Parameter guesses from the optimisation library are transparently 
passed to the client without restarting the virtual machine at each call. The client program inserts the guesses 
to the XML template and sends the description to a workflow submission servlet. The client waits until the workflow 
is processed and the value of the objective function is available on the web server as a result file. 
It returns the value to the calling function of the optimisation library.

Insertion of tasks and jobs in the engine tables and linkage of the dependencies 
are established by the submission servlet controlled by attributes in the XML workflow description. 
The attributes \ttt{start/closeTaskChain} and \ttt{start/closeTaskGroup} 
mark the beginning and end of task chains and groups. 
The attribute \ttt{startTaskChain} establishes a linear chain: 
every task except the very first is blocked by its direct predecessor (cf. section \ref{task_chains}).

The flag \ttt{startDependsOnGroupNode} indicates that all tasks of a chain have to be 
executed on the same node of a client group. It allows to distribute task chains to client groups 
where the client receiving the first task will also process all consecutive tasks of the chain. 
This allows more flexiblity than assigning all pending jobs to one single task, since client monitoring 
and job repetition can only be applied on a task-wide scale. The flag is stored in the column \ttt{DependsOnGroupNode}
of the \ttt{Tasks} table, and the servlet \ttt{TaskCompleted} updates the dependencies after 
a task from a chain has been completed:

\lstset{basicstyle=\ttfamily\fontsize{6.9}{6.9}\selectfont}
\lstset{language=SQL}
\begin{lstlisting} 
UPDATE Tasks SET Client='$client', ClientGroup=NULL
 WHERE DependsOnTask='$taskID' AND DependsOnGroupNode='true';
\end{lstlisting} 
Attribute \ttt{Type} annotates the types of data for the database table corresponding to the job's name.
If not in place, the workflow engine automatically constructs the table \ttt{JobReplaceTag} with 
columns \ttt{r1}, \ttt{r2}, \ttt{r3} of type \ttt{DOUBLE} for the parameter values and \ttt{Input}, \ttt{Output} of 
type \ttt{VARCHAR} for the names of the input or output files respectively.

The attribute \ttt{startTaskGroup} is used to split the workflow and create parallel branches. 
Each branch of a task group constitutes an independent chain, the initial tasks of all chains are coupled to the
same ID of a preceding task. This preceding task unblocks all branches simultaneously. 
Conversely \ttt{closeTaskGroup} acts as a thread barrier waiting until all tasks of a group are completed. 
It employes the column \ttt{TaskGroup} of the \ttt{Tasks} table and a dummy task as special job \ttt{MonitorTaskGroup} 
monitoring the status of all tasks with a certain group ID.
In the following example two independent branches (tasks 378 and 384) are already processed by 
nodes 04 and 02. Nodes 01 - 04 are now active on further tasks (cf. dotted lines):

\lstset{basicstyle=\ttfamily\fontsize{5.0}{5.0}\selectfont}
\lstset{language=sh}
\begin{lstlisting} 
select * from Tasks;
+-----+---------------+-----+--------+---------+---------+------+-------+---------+
| Task|               | Job | Dep's  |         | Client  | DepOn| Task  |         |
| ID  | Job           | ID  | OnTask | Client  | Group   | GrNod| Group | Status  |
+-----+---------------+-----+--------+---------+---------+------+-------+---------+
|.376.|.Simulation....|..56.|........|.node-03.|.cluster.|......|...375.|.active..|
| 377 | Upload        |  86 |    376 |         | cluster | true |   375 | waiting |
| 378 | Simulation    |  57 |        | node-04 | cluster |      |   375 | done    |
| 379 | Upload        |  87 |        | node-04 | cluster | true |   375 | done    |
|.380.|.Simulation....|..58.|........|.node-01.|.cluster.|......|...375.|.active..|
| 381 | Upload        |  88 |    380 |         | cluster | true |   375 | waiting |
|.382.|.Simulation....|..59.|........|.node-04.|.cluster.|......|...375.|.active..|
| 383 | Upload        |  89 |    382 |         | cluster | true |   375 | waiting |
| 384 | Simulation    |  60 |        | node-02 | cluster |      |   375 | done    |
| 385 | Upload        |  90 |        | node-02 | cluster | true |   375 | done    |
|.386.|.Simulation....|..61.|........|.node-02.|.cluster.|......|...375.|.active..|
| 387 | Upload        |  91 |    386 |         | cluster | true |   375 | waiting |
| 388 | Simulation    |  62 |        |         | cluster |      |   375 | waiting |
| 389 | Upload        |  92 |    388 |         | cluster | true |   375 | waiting |
|:390:|:MonitorTaskGr:|:375:|::::::::|:::::::::|:::::::::|::::::|:::::::|:active::|
| 391 | CreateReport  |  13 |    390 | server  |         |      |       | waiting |
+-----+---------------+-----+--------+---------+---------+------+-------+---------+
\end{lstlisting}
The monitor is implemented in the \ttt{Tasks} servlet. If the number of tasks 
with group ID 375 and status unequal to \ttt{done} turns zero, the servlet sets the monitor task to \ttt{done} 
and unblocks dependent tasks, here task 391 that generates a report of all performed simulation runs.

Using the workflow submission package we have interfaced the engine with the Adaptive Simulated Annealing
optimisation library \cite{asa1,asa2}. Simulated Annealing attempts to find the global minimum 
of the objective function in a stochastic process \cite{sa}. In Figures~\ref{fig:parameters} and~\ref{fig:modes} 
the black dots indicate the progress of the optimisation procedure. 
The final parameter set ${\rm \bf r^*} = (3.521, 3.755, 5.937)$~mm yields a plate with the exact frequency spectrum 
demanded. In the next section we analyse how robust the solution will be if the parameters differ from this 
optimal setting.

\subsection{Robustness of solutions \label{robust_design}}

The data from the previous sections is employed in a visualisation workflow analysing the robustness of a design 
against parameter fluctuations. The workflow automatically creates diagrams with the 
interactive data plotting utility Gnuplot \cite{gnuplot} and the R package for statistical computing \cite{R}. 
The parameter sets from the DOE study and the optimisation process are reused 
to interpolate the frequency response ${\rm \bf f}({\rm \bf r})$ around the optimum value ${\rm \bf r}^*$. 
Various simulated points in the immediate vicinity allow a detailed prediction for further parameter sets 
without additional numerical simulations.

\begin{figure*}
  \includegraphics[angle=-90,width=0.39\textwidth]{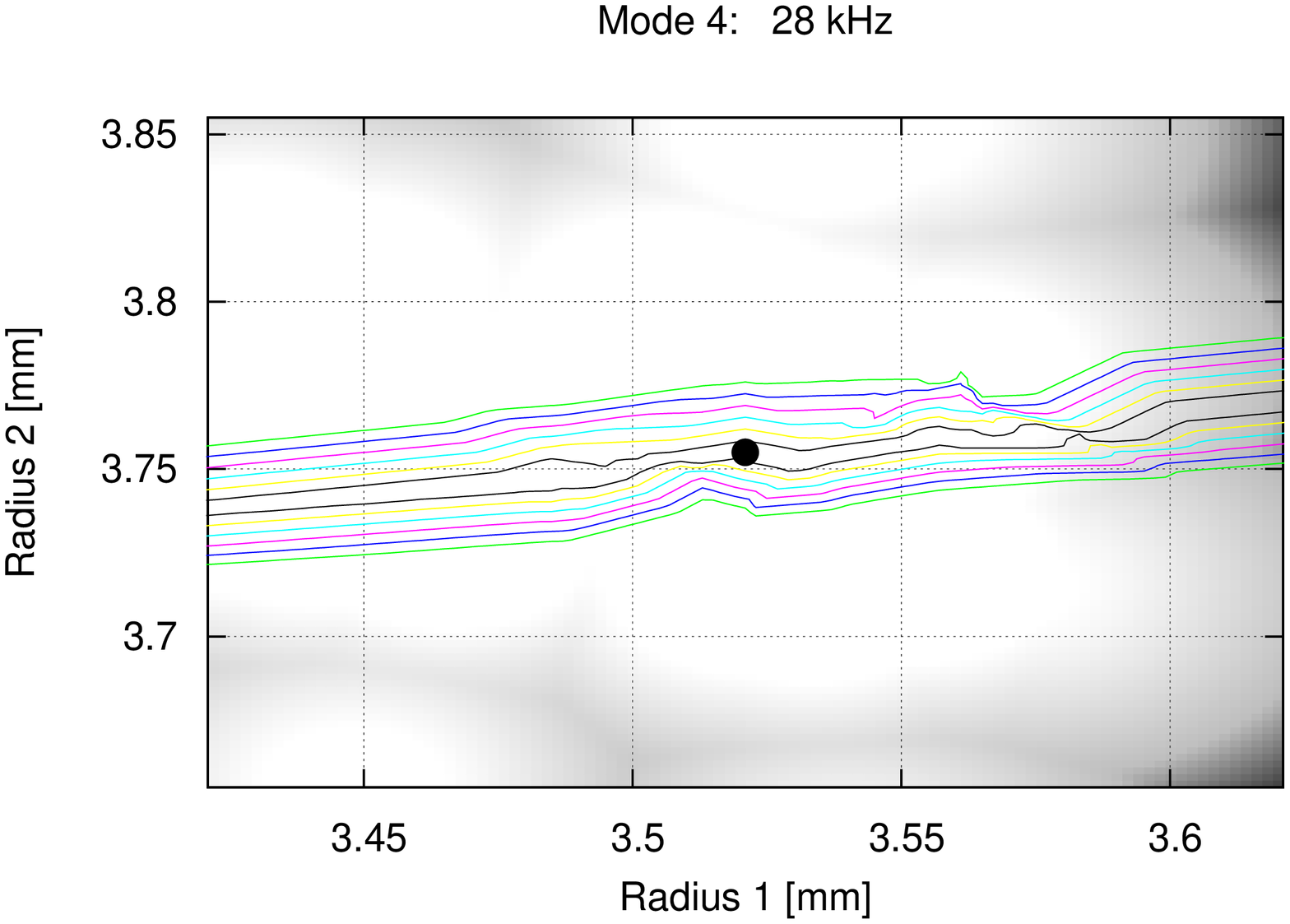}
  \hspace*{-1.9cm}
  \includegraphics[angle=-90,width=0.39\textwidth]{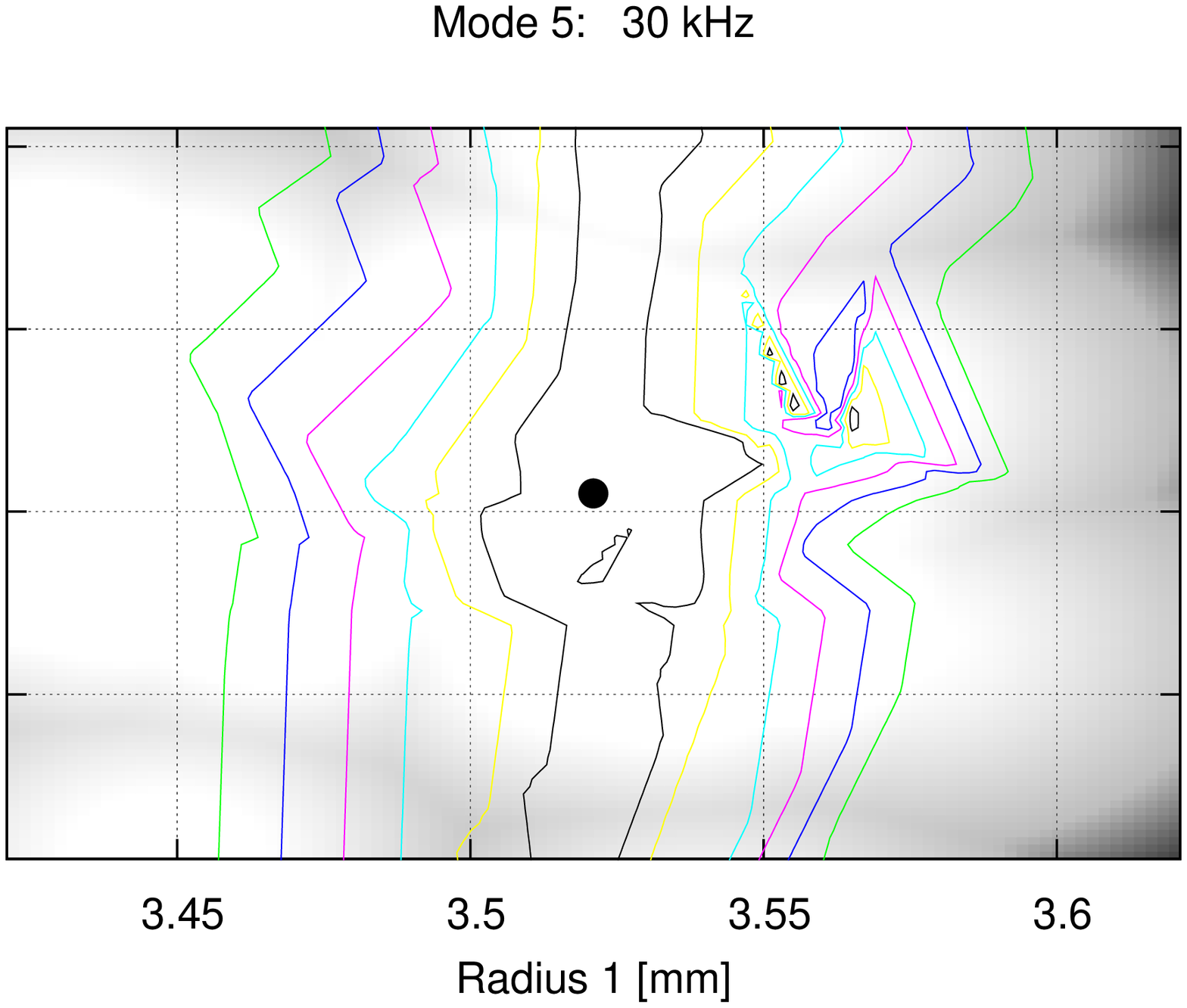}
  \hspace*{-1.9cm}
  \includegraphics[angle=-90,width=0.39\textwidth]{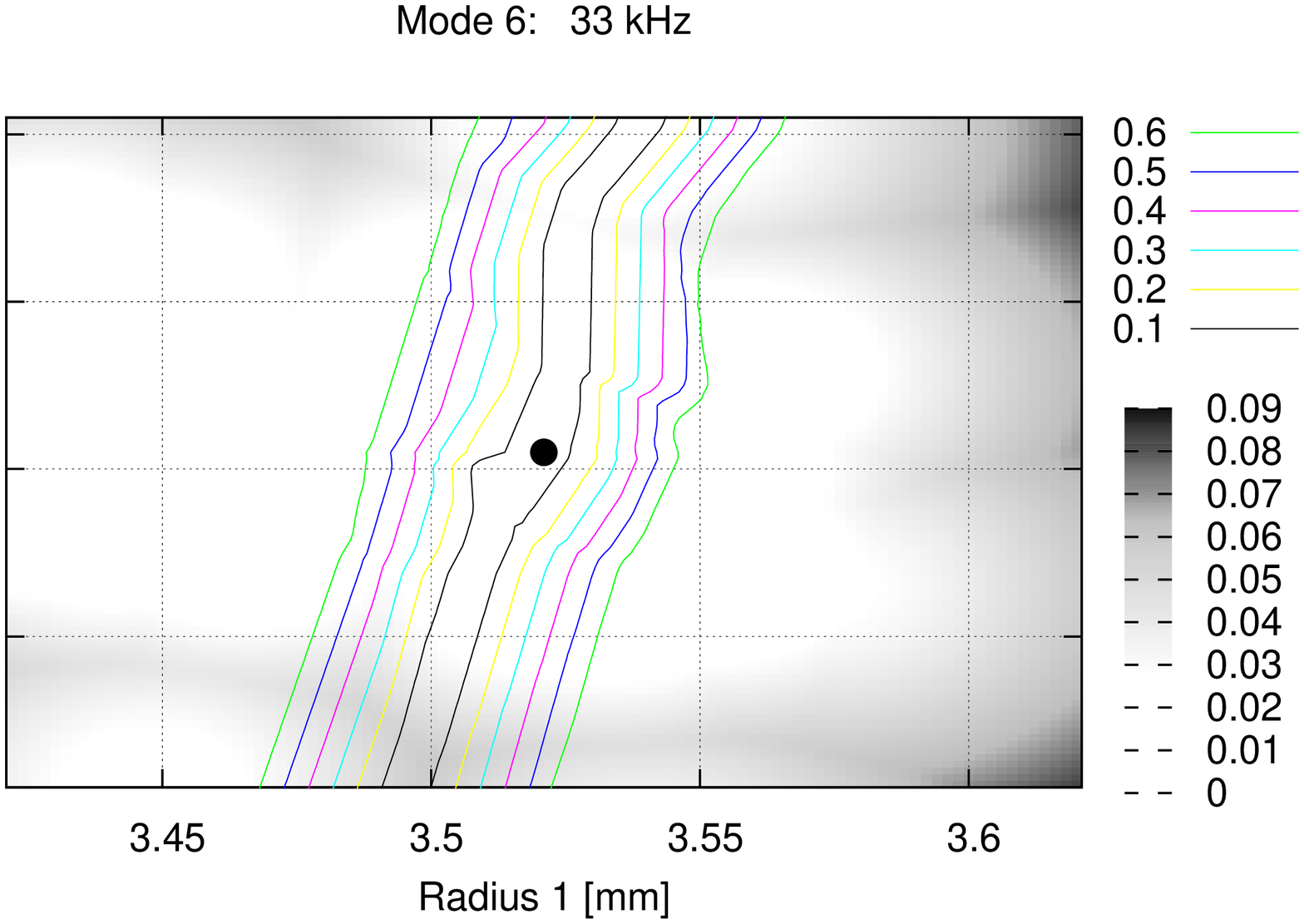}

  \vspace*{-1cm}

  \includegraphics[angle=-90,width=0.39\textwidth]{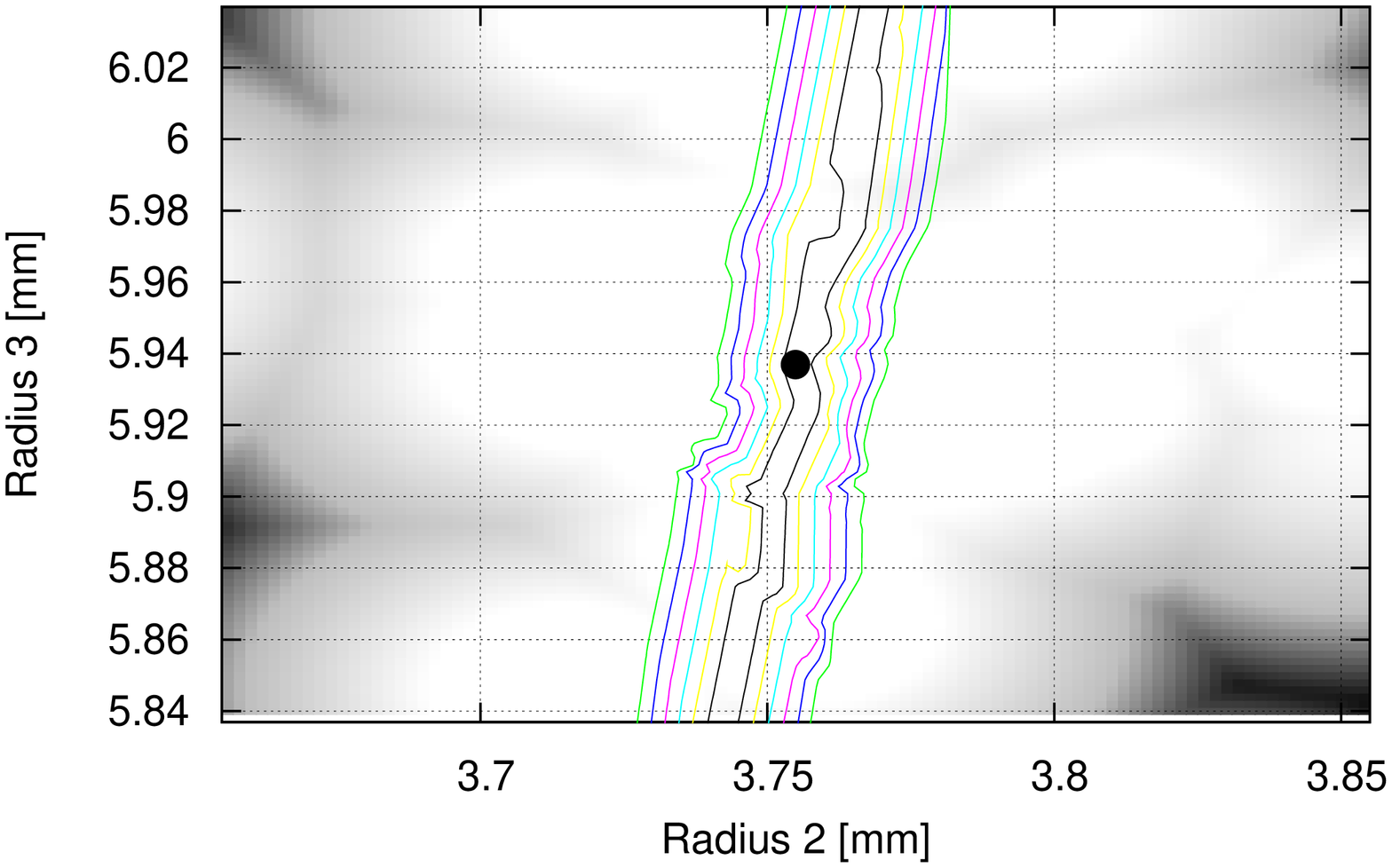}
  \hspace*{-1.9cm}
  \includegraphics[angle=-90,width=0.39\textwidth]{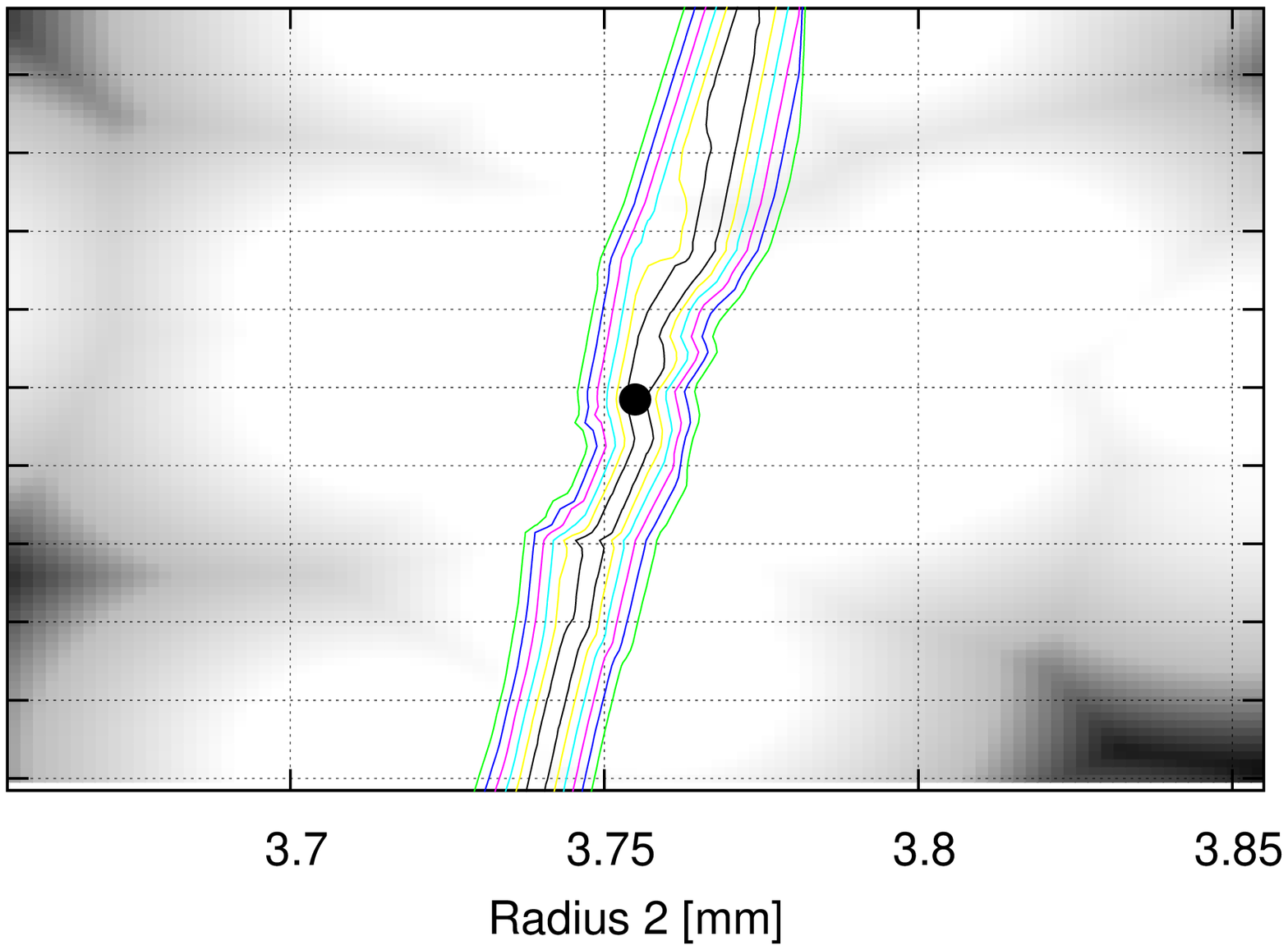}
  \hspace*{-1.9cm}
  \includegraphics[angle=-90,width=0.39\textwidth]{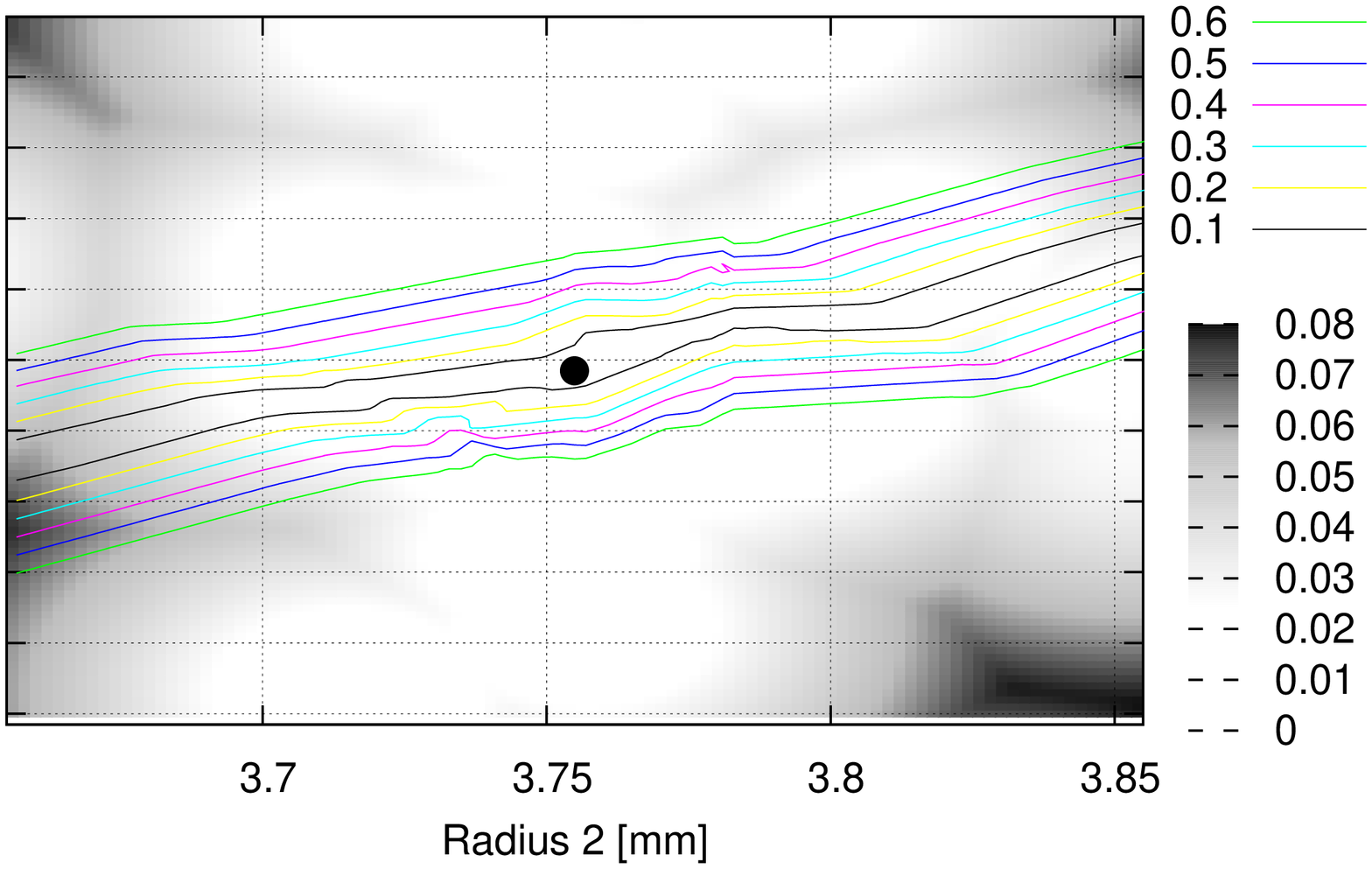}

  \vspace*{-1cm}

  \includegraphics[angle=-90,width=0.39\textwidth]{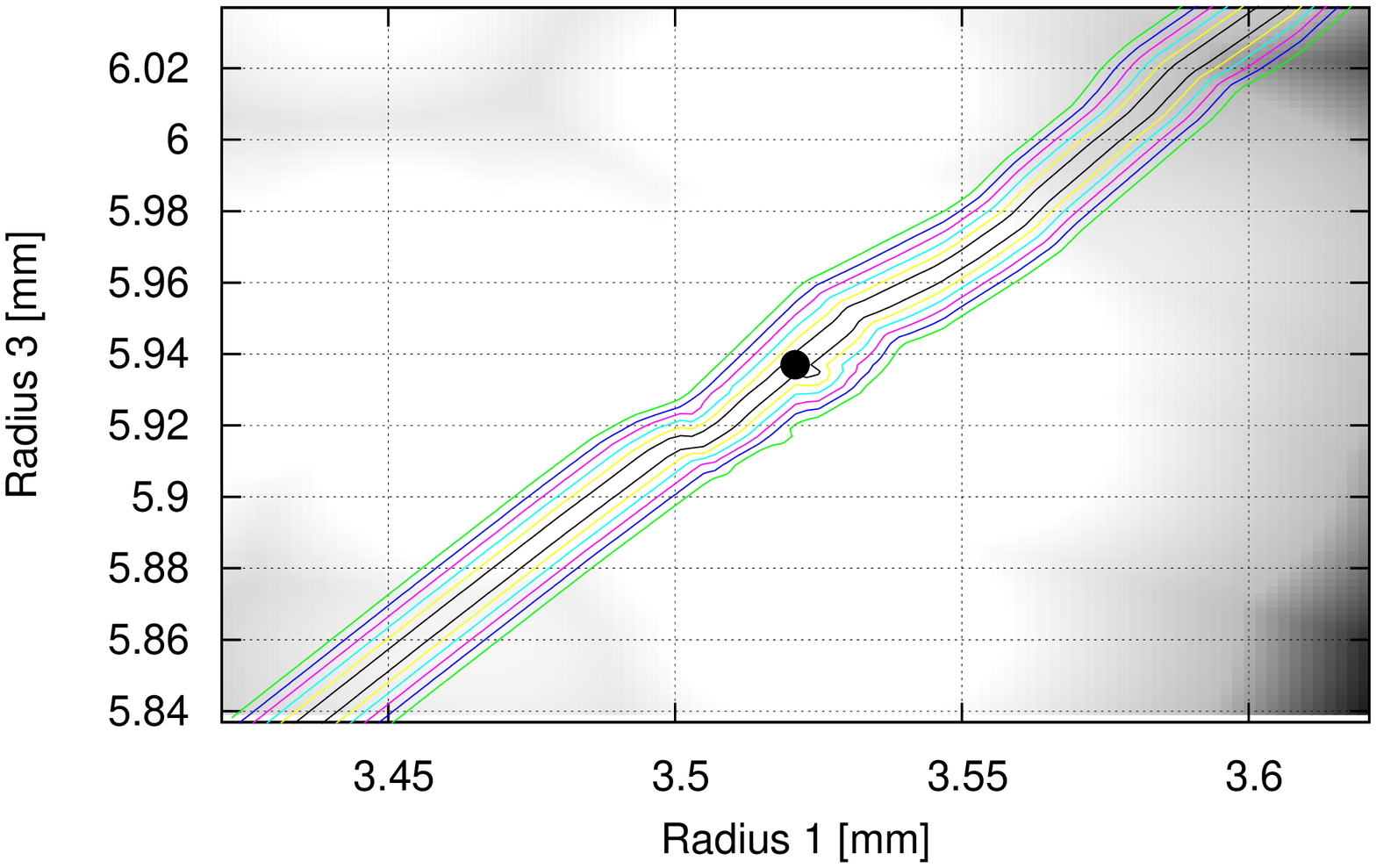}
  \hspace*{-1.9cm}
  \includegraphics[angle=-90,width=0.39\textwidth]{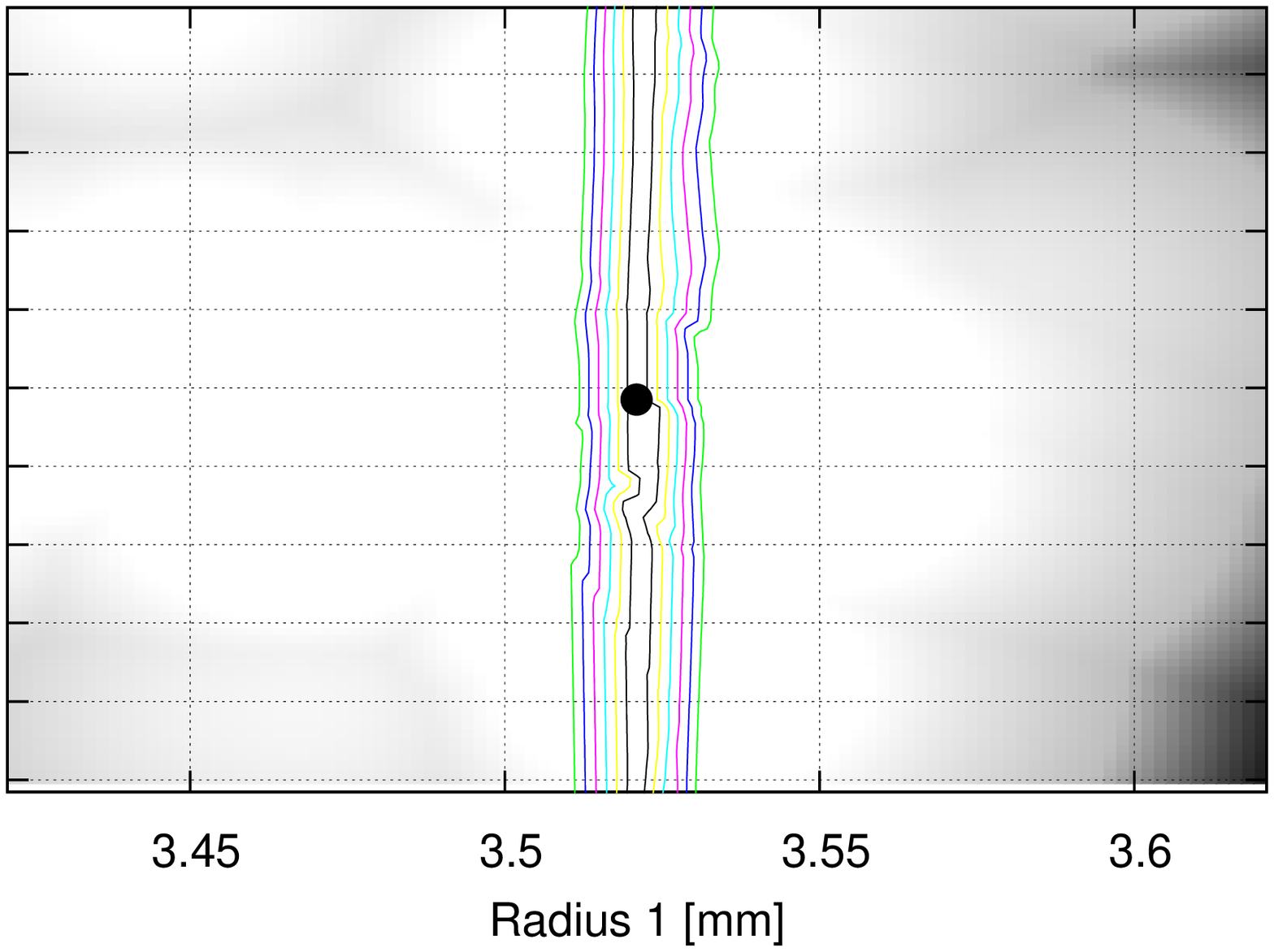}
  \hspace*{-1.9cm}
  \includegraphics[angle=-90,width=0.39\textwidth]{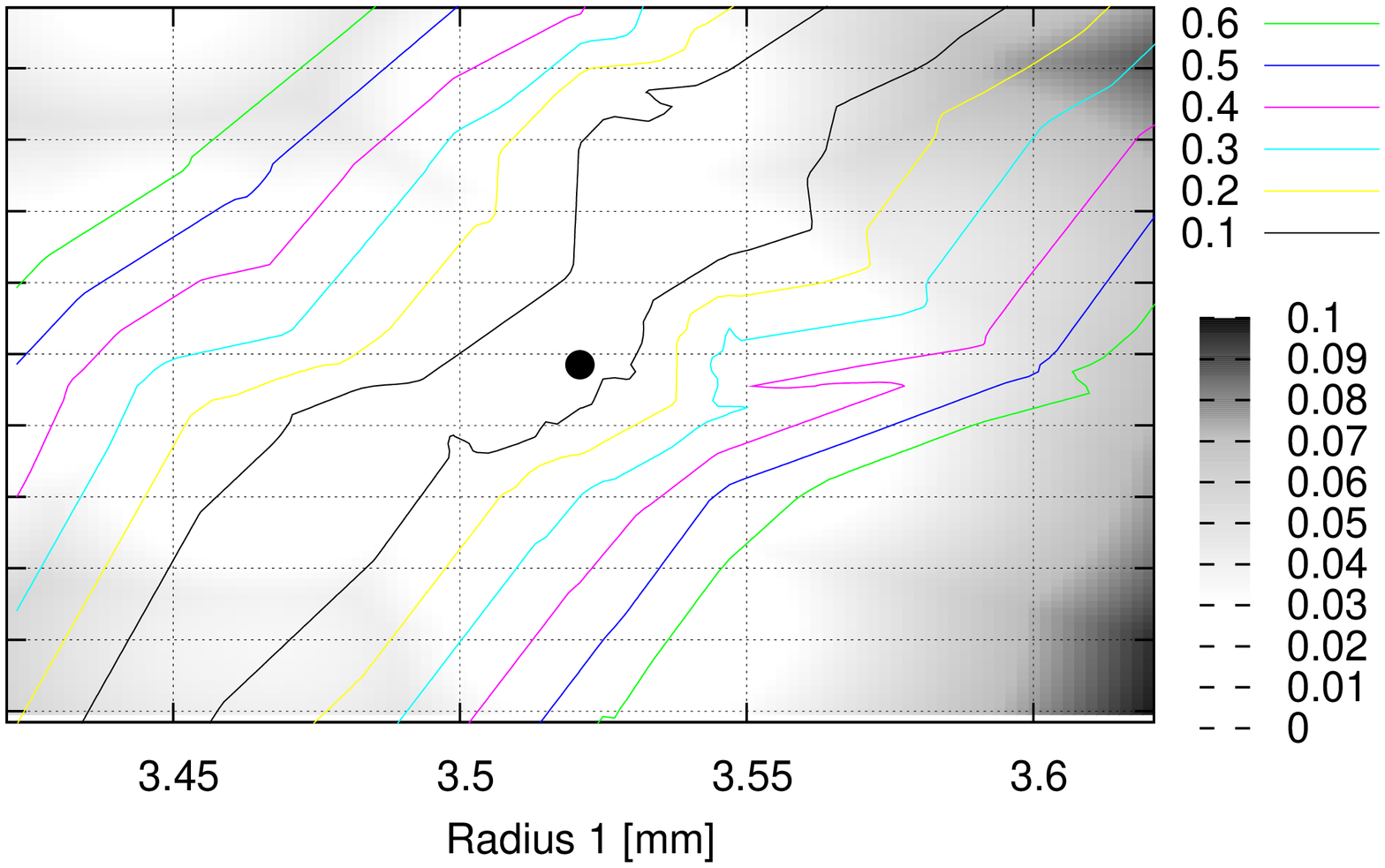}
  \vspace*{-0.3cm}	
  \caption{\label{fig:robust} Robustness of the solution: 
  Hole radii $r^*_1=3.521$, $r^*_2=3.755$ and $r^*_3=5.937$~mm (point in the center) reveal exactly 
  the desired mode frequencies.
  Contour lines indicate the frequency variation of deviations from this optimal point, denoted 
  in per mille of the original value, i.e. for mode 5 (30 kHz) the first line "0.1 ---" corresponds to 29.997 and 30.003 kHz. 
  The contour lines derive from the interpolation of the scattered simulation data. 
  Their reliability is heavily dependent of the distance to the closest simulation point
  in the parameter space. The distance from interpolated points to the nearest simulated point 
  is depicted as grey scale in the background in [mm]. In white areas a simulated parameter 
  set ($r_1$, $r_2$, $r_3$) is always within the close distance of maximal 0.03 mm.}
\end{figure*}

\begin{figure}
 \begin{center}
  \includegraphics[angle=-90,width=0.34\textwidth]{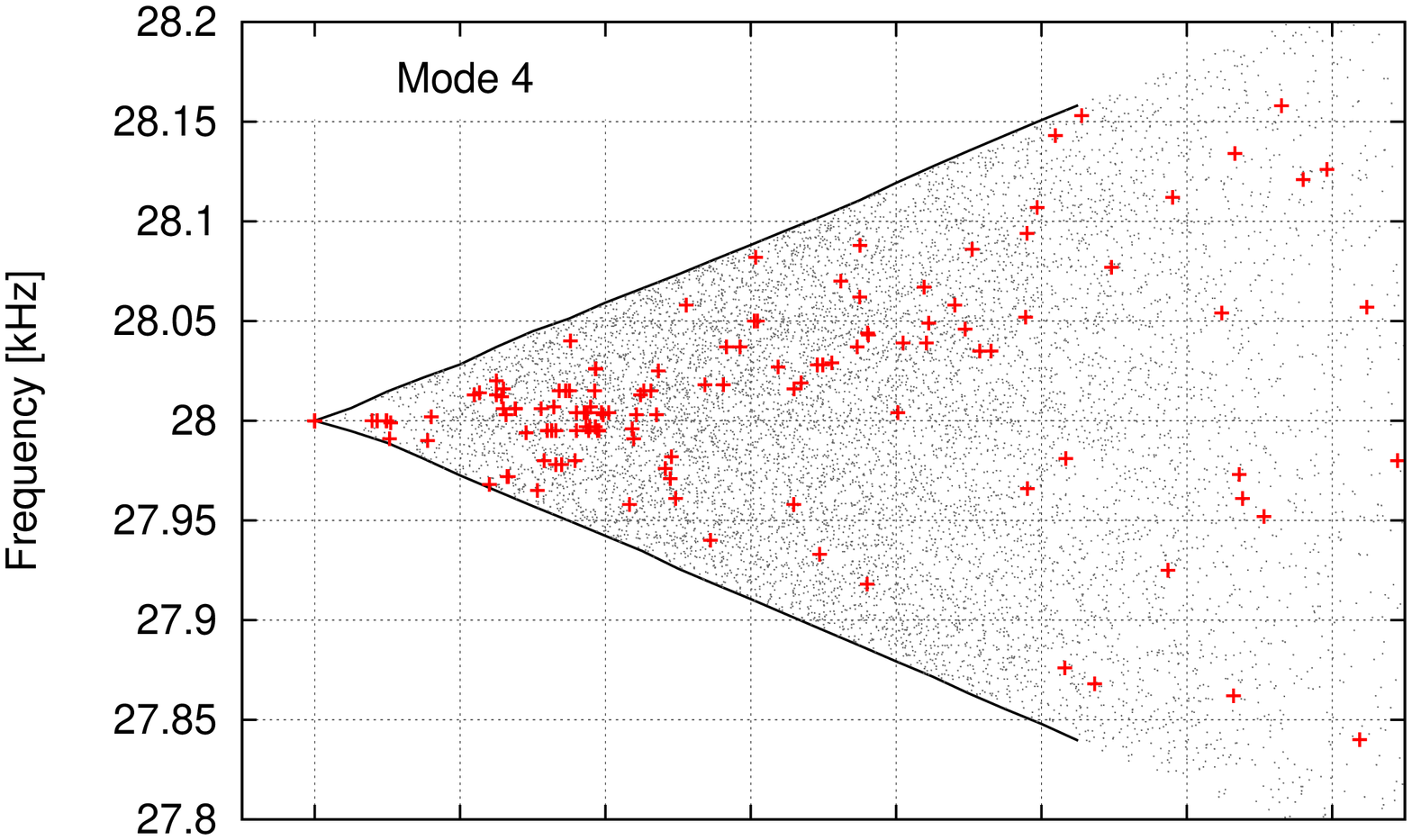}

  \vspace*{-0.7cm}
  \includegraphics[angle=-90,width=0.34\textwidth]{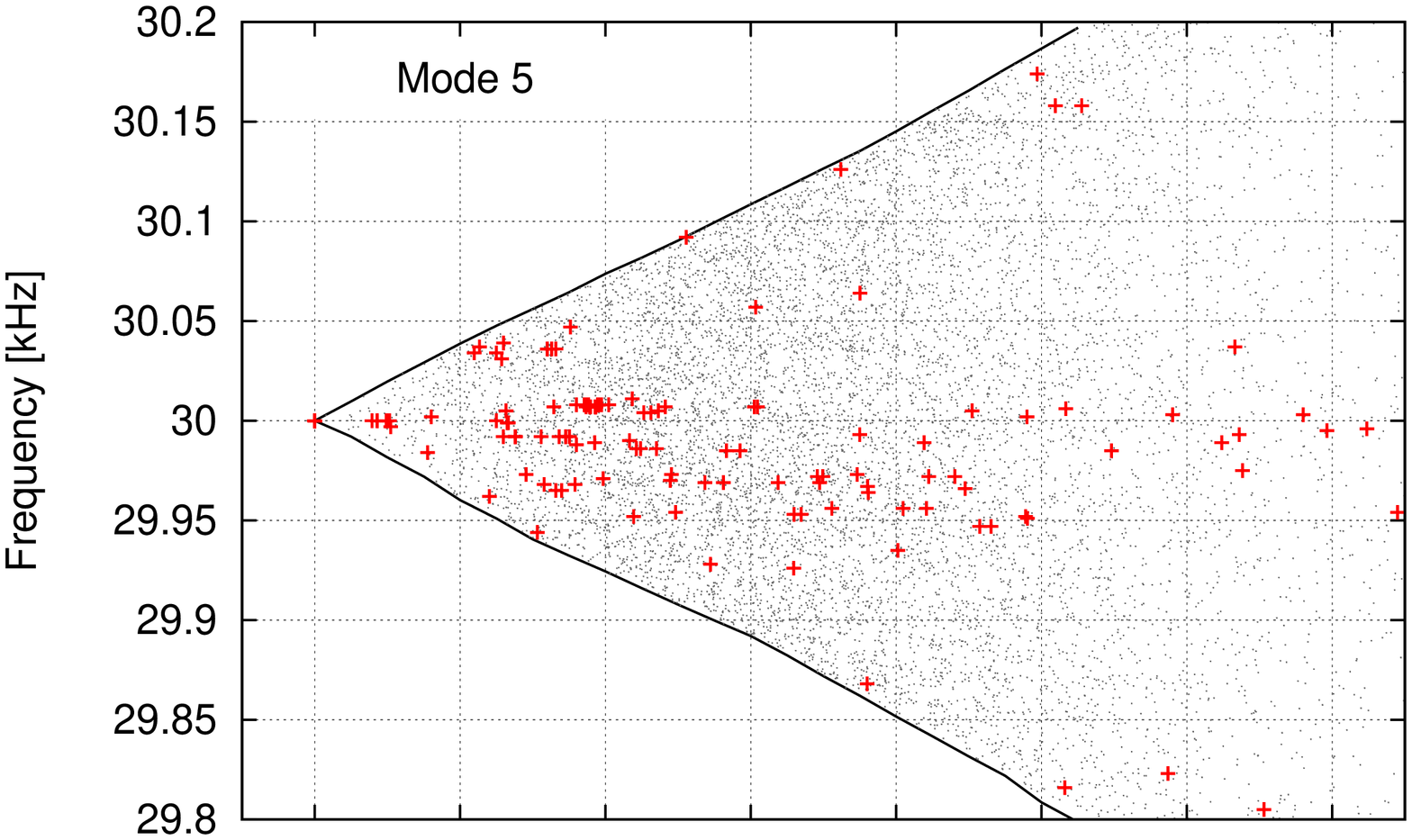}

  \vspace*{-0.7cm}
  \includegraphics[angle=-90,width=0.34\textwidth]{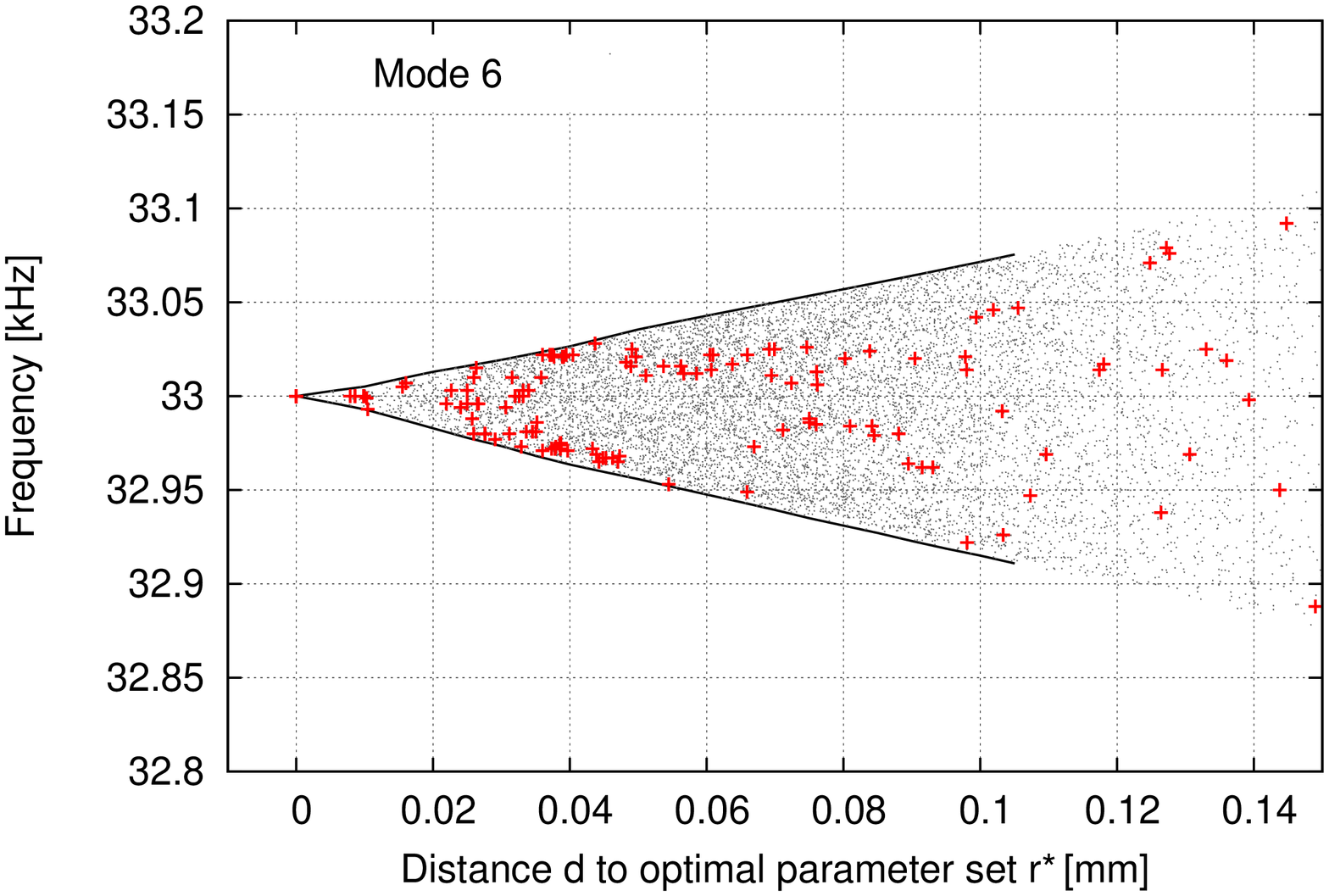}
\end{center}
  \caption{\label{fig:distances} Variation of mode frequency with parameters $(r_1, r_2, r_3) = {\rm \bf r}$ differing 
form their optimum value ${\rm \bf r^*}$. The Distance from the optimal value is defined by $d = \left[(r_1 \!-\! r^*_1)^2 + (r_2 \!-\! r^*_2)^2 + (r_3 \!-\! r^*_3)^2\right]^{1/2}$, red crosses mark 
data points from simulations runs, black dots are interpolated. Solid lines indicate worst expected values, i.e. the small aperture 
angle indicates that mode 6 is most robust against parameter variation.}
\end{figure}

\begin{figure}
  \includegraphics[angle=-90,width=0.172\textwidth]{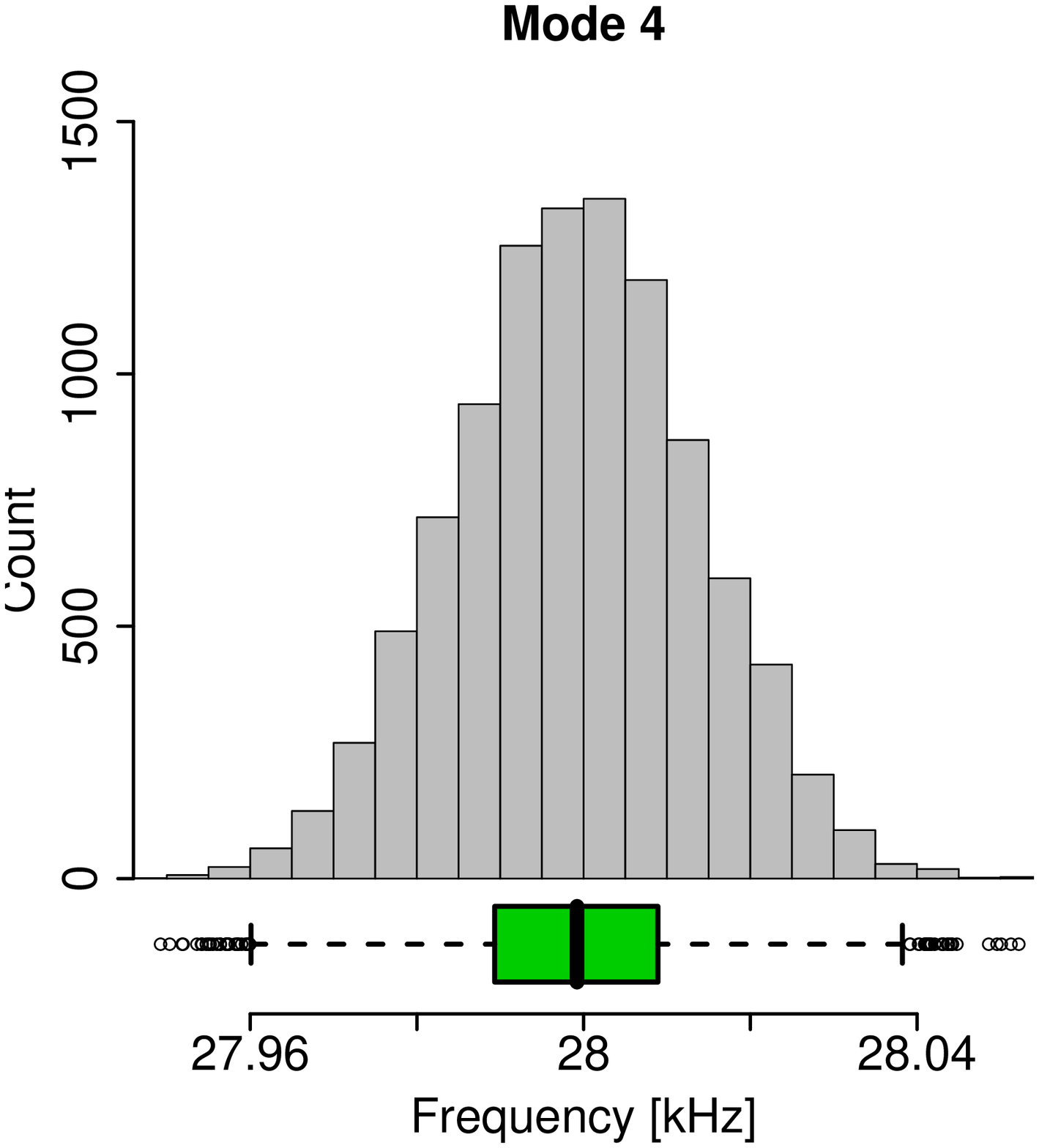}
  \hspace{-0.2cm}
  \includegraphics[angle=-90,width=0.172\textwidth]{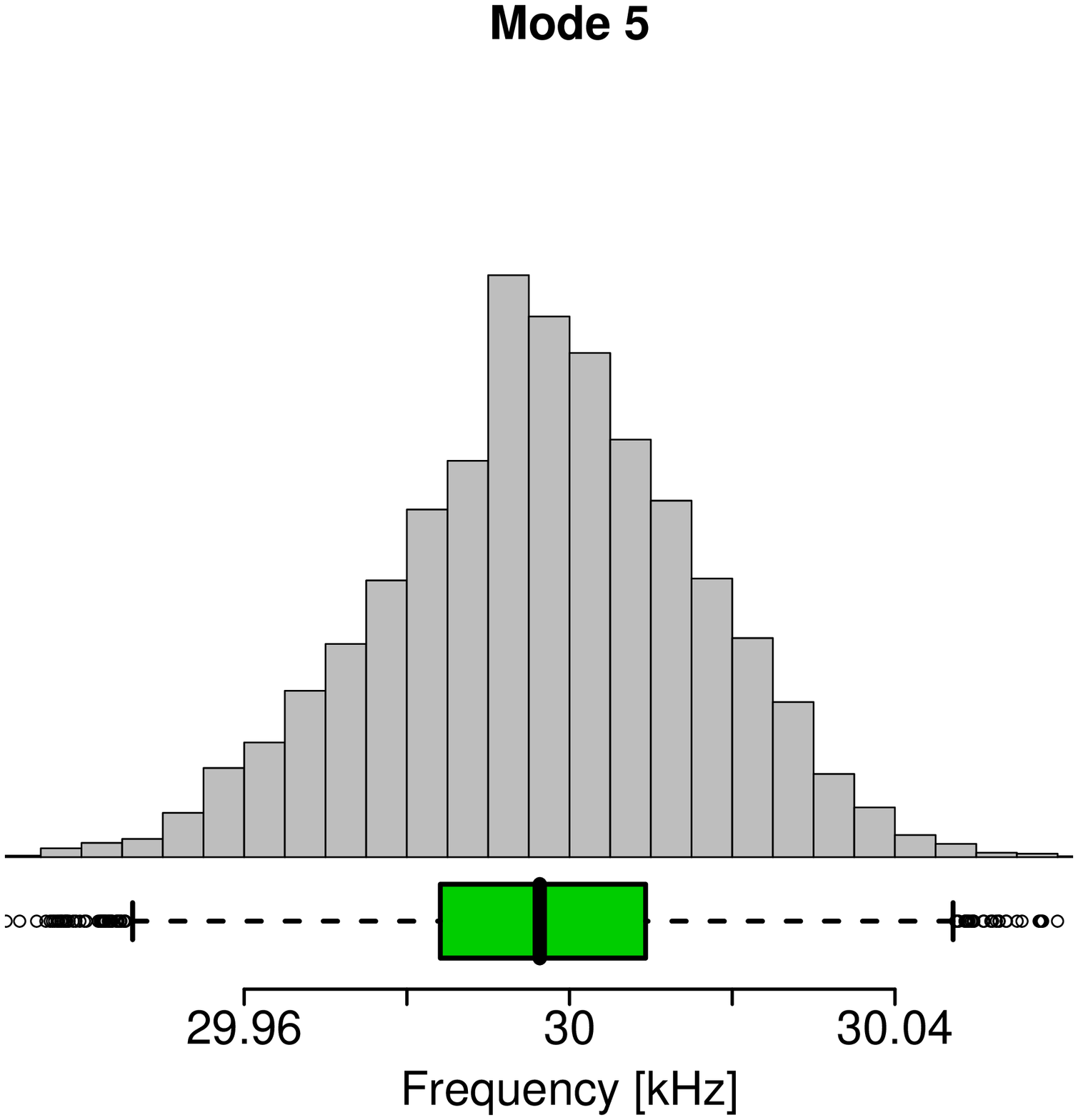}
  \hspace{-0.8cm}
  \includegraphics[angle=-90,width=0.172\textwidth]{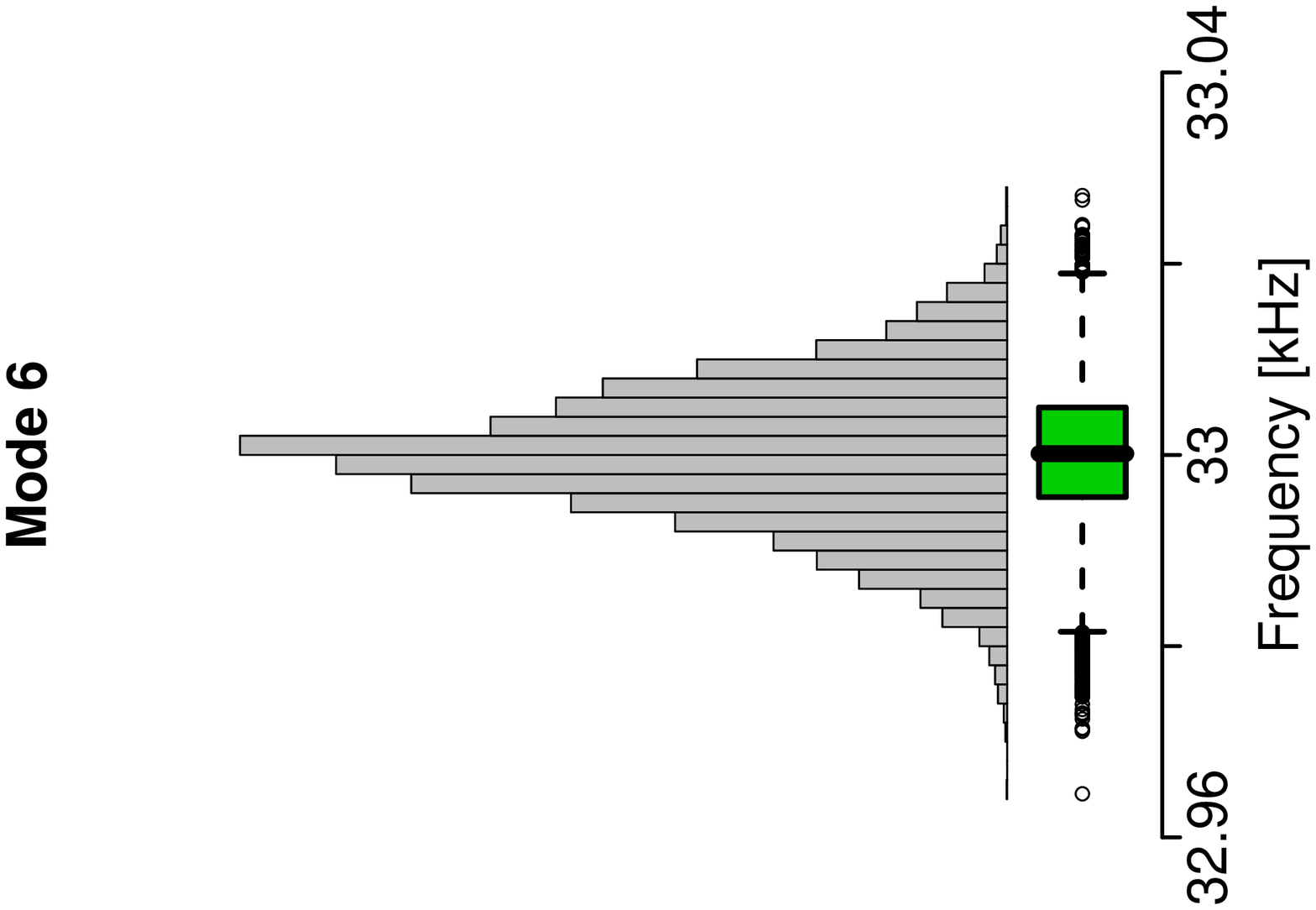}

  \caption{\label{fig:histogram}Distributions of natural frequencies if hole radii ${\rm \bf r}$ follow normal distributions 
around their optimal value ${\rm \bf r^*}$ with standard deviation $\sigma = 0.01$~mm, i.e. approximately 95$\%$ of 
the samples are within range ${\rm \bf r^*}\pm 2\sigma$. 10,000 interpolated simulations were used to create the histograms 
and the boxplots at the bottom, green boxes mark 50$\%$ quartiles for the results.} 
\end{figure}

A service supplied with a set of measured points and a list of parameters with unknown values executes the interpolation. 
The service tessellates the measured points using Qhull \cite{qhull,quickhull}, 
in the 3-dimensional parameter space, i.e. it constructs tetrahedrons between every four 
points with no further point inside. The values of arbitrary parameter points 
are then interpolated by the corner values of the tetrahedron containing the new point.
We have employed the interpolated frequency response to examine different aspects of the 
robustness of the optimal design:

\begin{itemize}
\item To get frequency contour lines in the parameter space as depicted in Figure~\ref{fig:robust} 
we have evaluated the frequency response function on regular grids in three perpendicular planes. 
The contour lines show how the value of the objective function varies for parameter sets next to 
the optimal point ${\rm \bf r^*}$ (black dot). Large distances between contour lines indicate areas 
of robust solutions with the frequency value being only slightly affected by parameter inaccuracies.
The reliability of this prediction depends on the distance between an interpolated parameter point ${\rm \bf r}$ and 
the next simulated point, cf. the grey background scale in Fig.~\ref{fig:robust}. 
Since part of the data stems from the optimisation run many simulated points are close to ${\rm \bf r^*}$. 

\item We have also interpolated the function on a cloud of random points ${\rm \bf r}$ around the optimum. 
Figure~\ref{fig:distances} shows the resulting frequencies depending on the 
distance $d=|{\rm \bf r}-{\rm \bf r^*}|$ to the optimal point ${\rm \bf r^*}$. 
Deviations from the optimal parameter set result in enlarged frequency spectra, the solid lines 
confine the spectra and specify worst case frequency values for parameter inaccuracies. 

\item Capability study techniques treat inevitable inaccuracies in the manufacturing processes by 
specific parameter distributions, e.g. the realisations ${\rm \bf r}$ are assumed to be normal 
distributed around their optimum value ${\rm \bf r^*}$. Histograms or box plots of the resulting 
frequency spectrum are used in Six Sigma analysis to determine how the manufacturing process meets 
specification limits \cite{sixsigma}.
\end{itemize}
For the last case we have generated 10,000 parameter sets ${\rm \bf r}=(r_1, r_2, r_3)$ following a multivariate normal 
probability density with mean ${\rm \bf r^*}$ and standard deviation $\sigma=0.01$~mm for all $D=3$ parameters: 
\begin{equation} \label{probdensity}
p({\rm \bf r}) = \frac{1}{(2\pi)^{D/2} \, \sigma} \, \exp\left[ -\frac{1}{2} \left(\frac{{\rm \bf r}-{\rm \bf r^*}}{\sigma}\right)^2\right]
\end{equation}
To deal with independent or correlated deviations ($\sigma_1$, $\sigma_2$, $\sigma_3$) a more general form of (\ref{probdensity}) 
can be applied, see \cite{pattern}. 
The results are summarised in Figure \ref{fig:histogram}. For the given manufacturing precision almost $95\%$ of the plates have natural frequencies in the range 
$f_{\rm Mode 4} = [27.973; 28.029]$, $f_{\rm Mode 5} = [29.964; 30.039]$, and $f_{\rm Mode 6} = [32.983; 33.013]$ kHz.

The diagrams are created by R and Gnuplot jobs in the workflow chain. We have manually assembled 
a prototype of each diagram and used it as a template for wrappers of the visualisation programs. 
Workflows then produce all diagrams supplying the scripts with alternating pairs of column numbers 
and the corresponding axis labels. The listing shows a wrapper for Gnuplot creating the 
contour plots from Figure~\ref{fig:robust}. The file \ttt{s1} contains the Gnuplot controls 
from the template, script \ttt{s2} starts Gnuplot with these controls:

\lstset{basicstyle=\ttfamily\fontsize{6.8}{6.8}\selectfont}
\lstset{language=Java}
\begin{lstlisting} 
public class JobGnuplot extends Job
{
  String inputfile;
  String outputfile;
  String colX;
  ...
  
  public void setParameters(Element parameter)
  {
    inputfile  = getAttribute(parameter, "Input"   );
    outputfile = getAttribute(parameter, "Output"  );
    colX       = getAttribute(parameter, "Column_X");
    ...
  }

  public void run()
  {
    StringBuffer t = new StringBuffer();
    t.append("set term post color solid enhanced 8" + "\n");
    t.append("set out   '" + outputfile + "'"       + "\n");
    t.append("set title '" + title      + "'"       + "\n");
    t.append("set grid"                             + "\n");
    t.append("unset surface"                        + "\n");
    t.append("set contour base"                     + "\n");
    t.append("set cntrparam levels increm 0.0, 0.1" + "\n");
    t.append("set view 0, 90"                       + "\n");
    t.append("set xlabel '" + xlabel + "'"          + "\n");
    t.append("set ylabel '" + ylabel + "'"          + "\n");
    t.append("set xrange[" + x1 + ":" + x2 + "]"    + "\n");
    t.append("set yrange[" + y1 + ":" + y2 + "]"    + "\n");
    t.append("splot '" + inputfile + "' using " + 
     colX + ":" + colY + ":" + colZ + " with lines" + "\n");
  
    String controlfile = "gnuplot.control";
    Script s1 = new Script(dir, controlfile);
    s1.createScript( t.toString() );

    String text = "GNUPLOT " + controlfile;
    Script s2 = new Script(dir, "start_gnuplot");
    s2.createScript(getShellHeader() + replaceAliases(text));
    s2.executeScript();

    flag_done = true;
  }
}
\end{lstlisting} 
Figures \ref{fig:robust} - \ref{fig:histogram} show the results of the visualisation workflows. 
In the diagrams only the influence of design parameter inaccuracies are examined. Other parameters, 
not subject to the design or optimisation procedure, may have more impact on the robustness of a solution, 
e.g. the thickness of the plate, homogeneity of the material, positions of the holes, 
or frequency shifts by temperature. Their influence can be evaluated in additional DOE studies applying 
the same methods and automated workflows discussed before.

%

\subsection{Further aspects}

\subsubsection{Stateful services}

The interpolation service from the previous section applies a list of simulated parameters with 
known frequencies as input. Every call of the service results in a new tessellation of all points 
from the list. This redundancy can be avoided by a stateful service, where the behaviour in consecutive calls
depends on the actual state of the service. In the case of the interpolation service, the tessellated models are stored 
in the local working directory. The MD5 algorithm \cite{md5sum} calculates hash keys for the content of the 
input files. To label the tessellated models keys with the length of 32 characters are used as file names. 
In every call the service first checks whether a model file for the input data exists. 
When identical input data occur, as many times during the generation of the diagrams, 
the service works with the already tessellated models from the repository. An implementation of the 
interpolation as a web service further increases the performance, 
since the model data can be held in the main memory of the application server.

%
%
%
%

\subsubsection{Billing}

As the runtime of all processed tasks is logged in the table \ttt{TasksCompleted} it can be used
to implement a billing system for expensive simulation tasks. In the database we have selected the 
runtimes of, e.g., all Ansys simulations and grouped them by users. The runtimes are multiplied with 
performance factors from the \ttt{Clients} table describing the individual speed of a node. 
If more processes than available processors (or cores) are active the runtimes are divided by 
load factor. The grid client monitors the load during the job execution and periodically sends 
it to the server. The time interval is customisable, the default is every two minutes. 
For the billing system a cron job logs the varying loads from table \ttt{ClientsAvailable} for further evaluations.

We developed standard benchmark jobs to automatically get presets for the performance factors,
e.g. we use a program to calculate the Mandelbrot set \cite{mandelbrot} for different resolutions and iteration depths. 
The benchmark jobs are executed on the clients at different load levels. The runtimes of the jobs serve as a relative measure 
for the preformance of the clients.

\subsubsection{Dedicated servers and distributed file systems}

In the basic set-up, the web server for file transfers, the servlet container for communication 
with the clients, and also the database engine are active on the same host.
Especially during the transmission of extensive data the web server imposes a bottleneck
and slows down the concurrent processes. The central file repository can be relocated to a 
dedicated web server to enhance the performance. In the default configuration the grid client contacts 
servlet container and web server at the same address.

The concept of a server file system on the engine also allows to apply more sophisticated methods for data 
exchange. The system was originally designed to deal with clients behind 
firewalls interacting only via HTTP with the central server. If the clients 
share common directories via network file systems, e.g. Samba \cite{smb}, NFS \cite{nfs}, 
or AFS with caching facilities \cite{afs,openafs}, the central file repository can be 
replaced by such a resource. In this scenario download and upload jobs are simplified 
to native methods that copy data between local directories. Even now the network file 
system is responsible to share the data between the clients, this change does not 
effect the abstract workflow descriptions.

Also peer-to-peer file systems, which are highly optimised for data throughput, can be embedded as a
transport layer, e.g. XtreemFS \cite{xtreemfs}. They organise data in chunks and distribute them to 
all network nodes. Access to remote data initiates parallel data transfers from different hosts. 
Since the data chunks are replicated on several locations, the system is robust against failures of single nodes. 
In each case the workflow engine remains responsible for the synchronisation of the data transfers. 
A consecutive download process of a client will not be initiated before the upload of the data 
from an other client has been finished.

\section{Conclusion}

The proposed platform independent client-server-system quickly turns a large number of 
heterogeneous computers into a flexible computing grid without expensive 
installations or administrator rights. The system is robust against unreliable network 
connections or firewall restrictions as the clients periodically contact the server 
engine via stateless HTTP. The engine utilises database tables to implement 
the logic of complex parallel workflows. It is based on a table for tasks and servlets to communicate with the 
clients. Further tables and servlets extend the modular system to execute customised
applications, platform independent services, workflow submission, and status information of the system itself. 
All tables' structure is self-explanatory. The data is processed and combined by SQL statements 
from the servlets. The XML workflow descriptions and the database tables provide generic interfaces 
to couple further applications, such as e.g. optimisation libraries with the system.
 
The engine distinguishes job execution from data transmission. Client applications only 
work on data in local directories while the system is responsible to transfer the data. 
Testing and debugging of complex workflows benefit from the fact that all intermediate 
data is accessible in the client's working directory. Workflow chains can be tested step-by-step 
as job wrappers for applications can be started manually using their \ttt{main()} method. 
Data transfer can be substituted by alternative methods, e.g. network file systems or distributed peer-to-peer networks.

The development of the engine was motivated by actual requirements of engineering and 
scientific computing. We have demonstrated how to accomplish design-of-experiment studies 
and parameter optimisations in finite element applications. We want to stress that pre- 
and post-processing and data visualisation routines can easily be integrated in the 
workflow chains. The hierarchical description allows to extend and recombine existing 
workflows. The data management is extremely simplified as all application parameters and 
results are available in the database tables of the engine.

Currently, we are working on a consolidated version of the system that we plan to release as open source project. 
Besides a consistent renaming of the classes and a complete refactoring of the code, we aim to integrate 
a database abstraction layer to simplify the integration of client applications, e.g. Hibernate/JPA \cite{jpa}. 
Hibernate encapsulates database-specific implementations and creates table schemes directly from job wrapper entities. 
In addition, we are developing a web interface to control the engine by browser. We plan to enhance 
data transfer for cluster computing, where clients are mutually connected and data transfer 
via a central server is an redundant detour.

\ifCLASSOPTIONcompsoc
  \section*{Acknowledgments}
\else
  \section*{Acknowledgment}
\fi

The authors would like to thank Yvonne Havertz from University of Cologne, 
Dr. Annette Kuhlmann from Shell Germany, and Dr. Detlef Billep from Fraunhofer Institute ENAS.

\bibliographystyle{IEEEtran}
\bibliography{references}

\end{document}